%% file: bare_jrnl.tex
\documentclass[journal]{IEEEtran}

\ifCLASSINFOpdf
\else
\fi
\include{pack}
\usepackage[hidelinks,draft=false]{hyperref}
\newcommand{\sm}[1]{\textcolor{black}{{#1}}}

\begin{document}

\title{Neural Estimators for Conditional Mutual Information Using Nearest Neighbors Sampling}

\author{Sina~Molavipour,~\IEEEmembership{Student~Member,~IEEE,}
	    Germ\'{a}n~Bassi,~\IEEEmembership{Member,~IEEE,}
	    and~Mikael~Skoglund,~\IEEEmembership{Fellow,~IEEE}
    	\thanks{The authors are with the school of Electrical Engineering and Computer Science, KTH Royal Institute of Technology, Stockholm, Sweden 100 44. (e-mails: \{\tt sinmo, germanb, skoglund\}{\tt @kth.se})}
   		\thanks{This work was supported in part by the Knut and Alice Wallenberg Foundation and the Swedish Foundation for Strategic Research.}
}

\maketitle

\begin{abstract}
The estimation of mutual information (MI) or conditional mutual information (CMI) from a set of samples is a long-standing problem.
A recent line of work in this area has leveraged the approximation power of artificial neural networks and has shown improvements over conventional methods.
One important challenge in this new approach is the need to obtain, given the original dataset, a different set where the samples are distributed according to a specific product density function.
This is particularly challenging when estimating CMI.

In this paper, we introduce a new technique, based on $k$ nearest neighbors ($k$-NN), to perform the resampling and derive high-confidence concentration bounds for the sample average.
Then the technique is employed to train a neural network classifier and the CMI is estimated accordingly.
We propose three estimators using this technique and prove their consistency, make a comparison between them and similar approaches in the literature, and experimentally show improvements in estimating the CMI in terms of accuracy and variance of the estimators.
\end{abstract}


\IEEEpeerreviewmaketitle

\section{Introduction}

\IEEEPARstart{C}{onditional} mutual information is recognized as an important statistical metric since, for example, characterizes the capacity of communication channels such as channels with random state and the relay channel~\cite{el2011network}; however, its relevance goes beyond communication scenarios.
Directed information~\cite{massey1990causality}, which is a notion for quantifying causal impact in stochastic processes, is computed as a possible infinite sum of CMIs~\cite{Mol2017test}. 
Additionally, CMI has been adopted in machine learning~\cite{fleuret2004fast, loeckx2009nonrigid} as a way to extract shared information in data, while in the information bottleneck method, it can be used as a regularizer~\cite{mukherjee2019machine,rodriguez2020variational}.

The estimation of information-theoretic quantities has been an important subject in statistical inference for many years. 
In general, conventional methods are categorized as parametric and non-parametric estimators.
In~\cite{wang2009universal} several of these methods for estimating entropy, mutual information, and relative entropy are reviewed.
One well-known non-parametric method to estimate MI of continuous random variables is the KSG estimator~\cite{kraskov2004estimating, gao2018demystifying}; this estimator is based on the $k$ nearest neighbors method ($k$-NN) and shows a favorable performance for data with small dimensions.
This method has subsequently been extended to estimate CMI in~\cite{runge2018conditional, vejmelka2008inferring, frenzel2007partial}.
However, observations show that as the dimension of the data increases, the estimation accuracy deteriorates, and addressing this issue has remained a challenge.

Recent studies leverage the power of artificial neural networks to improve the estimation of information-theoretic quantities.
In the recent work~\cite{belghazi2018mine}, the authors propose the use of neural networks to estimate MI and, according to their numerical experiments, promising improvements with respect to the conventional KSG method can be seen for high-dimensional data.
The key idea in~\cite{belghazi2018mine} is to estimate a lower bound for the MI---known as a variational bound---instead of directly estimating the MI; the network is trained to maximize this lower bound which results in a tight approximation of the MI.
This approach has been followed by a series of other works such as \cite{mcallester2018formal, poole2018variational, mukherjee2019ccmi_conf, song2019understanding,qin2019rethinking, molavipour2020conditional}. 
In particular, in~\cite{mcallester2018formal}, the limits of estimation using variational bounds are investigated and the authors provide high confidence bounds for these constraints in terms of the number of samples.
Similar arguments can be found in~\cite{song2019understanding} where the authors address the bias--variance trade-off in the neural estimators for MI.
A thorough comparison for MI estimators is done in~\cite{poole2018variational} and different methods based on variational bounds are compared in terms of bias and variance.
In~\cite{qin2019rethinking}, further insights and applications are provided for a neural estimation of MI.

Before proceeding, consider the definition of CMI for continuous random variables:
\begin{align}
I(X;Y|Z) &\coloneqq \iiint p(x,y,z)\log \frac{p(x,y,z)}{p(x|z)p(y,z)}dx\, dy\, dz \nonumber\\
&= \ex_{p(y,z)}\! \left[ \klD{p(x|Y,Z)}{p(x|Z)} \right].
\label{eq:CMI_def}
\end{align} 
A lower bound on the CMI can thus be obtained employing the Donsker--Varadhan (DV) variational characterization of the divergence~\cite{donsker1975asymptotic}:
\begin{multline}
I(X;Y|Z) \geq \ex_{\djoint} \big[ f(x,y,z) \big]\\
-\log \ex_{\dprod} \big[ \exp f(x,y,z) \big],
\label{eq:LB_DV}
\end{multline}
where $f(\cdot)$ is any function such that the two expectations exist and are finite.
The lower bound~\eqref{eq:LB_DV} may be relaxed, as suggested by Nguyen, Wainwright, and Jordan (NWJ) in~\cite{nguyen2010estimating}, resulting in the following lower bound:
\begin{multline}
I(X;Y|Z)\geq \ex_{\djoint} \big[f(x,y,z)\big]\\
- e^{-1}\ex_{\dprod}\big[ \exp f(x,y,z) \big].
\label{eq:LB_NWJ}
\end{multline}
These bounds are tight with the appropriate choice of $f(\cdot)$, and equality holds in~\eqref{eq:LB_DV} by choosing $f(\cdot)$ as
\begin{align}
f^*_{\DV}(x,y,z)\coloneqq C + \log \frac{p(x,y,z)}{p(x|z)p(y,z)}, \ \forall C\in\mathbb{R},\label{eq:optimal_f_DV}
\end{align}
while 
\begin{align}
f^*_{\NWJ}(x,y,z) \coloneqq 1+\log \frac{p(x,y,z)}{p(x|z)p(y,z)}\label{eq:optimal_f_NWJ}
\end{align} 
yields equality in both bounds~\eqref{eq:LB_DV} and~\eqref{eq:LB_NWJ}.
If the joint probability density function $p(x,y,z)$ were known, it would be possible to compute the optimal functions~\eqref{eq:optimal_f_DV} and~\eqref{eq:optimal_f_NWJ}, and respectively the bounds~\eqref{eq:LB_DV} and~\eqref{eq:LB_NWJ}. Most importantly, we could derive the CMI directly:
\begin{align}\label{eq:CMI_LDR}
I(X;Y|Z)= \ex_{\djoint}\left[f^*_{\LDR}(x,y,z)\right],
\end{align}
where
\begin{align}\label{eq:optim_f_LDR}
f^*_{\LDR}(x,y,z) \coloneqq \log \frac{p(x,y,z)}{p(x|z)p(y,z)}
\end{align} 
is the logarithm of the density ratio (LDR).
However, we only have access to a set of samples distributed according to $p(x,y,z)$.
Using these samples, we will approximate the functions~\eqref{eq:optimal_f_DV}, \eqref{eq:optimal_f_NWJ}, and~\eqref{eq:optim_f_LDR}, which will allow us to estimate the CMI according to~\eqref{eq:LB_DV}, \eqref{eq:LB_NWJ}, or~\eqref{eq:CMI_LDR}.

We note that, for any fixed function $f(\cdot)$, the NWJ bound~\eqref{eq:LB_NWJ} is looser than the DV bound~\eqref{eq:LB_DV} except for the case of~\eqref{eq:optimal_f_NWJ}; however, the former bound has the advantage of having a linear form, which may be useful when the bound is estimated empirically. 
As noted in~\cite{molavipour2020conditional}, the average of several estimates of the DV bound is neither a lower bound nor an upper bound of the CMI due to the concavity of the $\log(\cdot)$ function and Jensen's inequality.
This becomes of paramount importance if the estimation must not exceed the true value of the CMI. 
For instance, when estimating the capacity of a communication channel determined by a CMI, the estimated value must be below the true value of the CMI to ensure a reliable communication.
It is worth noting that, although estimating with insufficient number of samples may also cause such violation, this should not be confused with the issue caused by the non-linearity of the terms.
Nonetheless, if there is no constraint in the estimated value of the CMI being below the true value, we may safely use any of the aforementioned three estimators. 
In fact, we show in our experiments that, in some cases, estimations based on~\eqref{eq:CMI_LDR} are more accurate while being above the true value of CMI.

As previously mentioned, the authors of~\cite{belghazi2018mine} introduced the idea of using artificial neural networks to estimate MI; in particular, they calculate the DV bound, where $f(\cdot)$ is substituted with a neural network and the right-hand side (RHS) of~\eqref{eq:LB_DV} is maximized with the gradient descent method. 
A new approach to estimate both the MI and the CMI is taken in~\cite{mukherjee2019ccmi_conf}, where a neural network classifier is first trained to distinguish whether samples are generated according to the joint or product density function.
Then the authors show that the output of this classifier can be used to approximate the optimal functions in~\eqref{eq:optimal_f_DV} and~\eqref{eq:optimal_f_NWJ}.
However, instead of estimating the lower bounds on the CMI directly, they express the CMI as a difference of two MI terms, i.e., 
\begin{equation}
I(X;Y|Z) = I(X;Y,Z) -I(X;Z), \label{eq:CMI_2_MI}
\end{equation}
and estimate the DV (or NWJ) lower bound for each term separately.

In this paper, we adopt the classifier technique of~\cite{mukherjee2019ccmi_conf} and introduce a new method to apply it directly to the estimation of CMI. 
Estimating CMI is more complicated than estimating MI since the technique relies on having samples that are distributed according to the product density $p(x|z)p(y,z)$ apart from the original samples distributed according to $p(x,y,z)$.
The approach of~\cite{mukherjee2019ccmi_conf}, which estimates the two terms on the RHS of~\eqref{eq:CMI_2_MI}, only requires samples distributed according to $p(x)p(y,z)$ and $p(x)p(z)$, which are simple to obtain given the original samples.
Here, we address this issue in Section~\ref{sec:prelim} and show that the $k$-NN method can be employed to obtain the desired samples from the original data.
In fact, this technique can be applied to any resampling problem where we want to enforce a more restrictive factorization for the density function of the new samples. 
In Section~\ref{sec:main}, we establish concentration bounds for the empirical average with respect to data sampled according to our $k$-NN method, which is one of the main contributions of this paper. 
Next, the consistency of our proposed estimators is investigated by the approximation and generalization power of our setup.
Experiments and simulation results are presented in Section~\ref{sec:simul}.
Finally, we conclude the paper in Section~\ref{sec:conclude} where we discuss possible future direction.

\section{Preliminaries and challenges}
\label{sec:prelim}

Consider a dataset of $n$ triples $(X,Y,Z)\in \mathcal{X}^3$ where $X$, $Y$ and $Z$  are mappings $\Omega\to\mathcal{X}\subset \mathbb{R}^d$ with finite Lebesgue measure $\lambda(\mathcal{X})$.
For simplicity, we assume the mappings have the same range, while the extension is straightforward when variables range over different sets.
Each triple is generated i.i.d. according to $\djoint$. 
The classifier technique, which is used at the core of our estimators, relies on a binary neural classifier that distinguishes whether an input sample $(x,y,z)$ is more likely to be generated from the joint density $\djoint$ or the product density $\dprod$.

As neither of these density functions is known, estimating the CMI based on~\eqref{eq:LB_DV}, \eqref{eq:LB_NWJ}, or~\eqref{eq:CMI_LDR} encounters the following two challenges:
\begin{enumerate}
	\item The optimal functions $f_\DV^*$, $f_\NWJ^*$, and $f^*_\LDR$ cannot be computed due to the unknown densities, and thus must be approximated.
	
	\item Even if the previous point is solved, it is not possible to derive~\eqref{eq:LB_DV}, \eqref{eq:LB_NWJ}, or~\eqref{eq:CMI_LDR} analytically, and thus the expectations must also be approximated using the samples.
\end{enumerate}

In the following, we address these issues.
We will see that the output of the binary neural classifier, with a proper loss function, can be used to solve the first challenge.
However, this leads to a new problem; in order to train the neural classifier, we need samples distributed according to both the joint and the product density functions.
The solution to this new issue, which also addresses the second challenge, is to generate sample batches according to $\djoint$ and $\dprod$, where providing the latter is not straightforward and is the main focus of this paper.

\emph{Notation:}
Throughout the paper, capital letters (e.g., $X$) mostly denote random variables, while their lower-case counterparts (e.g., $x$) denote instances of said random variables.
We use the notation $x^n$ to denote the sequence of $x_1, \dots, x_n$.
However, we may also use $n$ in the superscript to emphasize the dependence on a quantity with $n$; this will be clear in the context.
Additionally, for an arbitrary set $\mathcal{I}$, $x_{\{1,\dots ,n\}\setminus \mathcal{I}}$ indicates the sequence of $x_i$'s, where $i$ iterates on $1,\dots ,n$ excluding the elements in $\mathcal{I}$.

\subsection{Resampling}

In this section, we explain how to generate the said batches of samples from the dataset $\{(X_i, Y_i, Z_i)\}_{i=1}^n$.
Define $\mathcal{I}_b$ to be a set of $b$ numbers picked uniformly at random (without replacement) from the set $\{1,\dots ,n\}$.
Let $\mathcal{B}_\textnormal{joint}^b$ denote the joint batch, which consists of $b$ samples distributed i.i.d. according to $\djoint$ and it is defined as:
\begin{align}
\mathcal{B}_\textnormal{joint}^b \coloneqq \big\{(X_i,Y_i,Z_i) \mid i\in \mathcal{I}_b \big\}.
\label{eq:jointBatch}
\end{align}
On the other hand, let $\mathcal{B}_\textnormal{prod}^{b'}$ be the product batch such that it contains $b'$ samples distributed according to $\dprod$.
To construct this batch, we exploit the notion of $k$ nearest neighbors ($k$-NN).

\begin{definition}
	\label{def:isoKnn}
	Assume the dataset $\{(x_i, y_i, z_i)\}_{i=1}^n$ of size $n$ is given.
	Let $\mathcal{I}_{m}$ be a set of $m$ indices chosen uniformly at random without replacement from $\{1,\dots,n\}$, and $\mathcal{I}^c_{m}\coloneqq \{1,\dots,n\}\setminus \mathcal{I}_m$.
	For any $\zeta\in\mathcal{X}$, define $\mathcal{A}^{m,k,n}(\zeta,z^n)$ as the set of indices of the $k$ nearest neighbors of $\zeta$ (by Euclidean distance) among $z_i$, for $i\in \mathcal{I}^c_{m}$.
	In other words, 
	let $\pi_{\zeta}:\{1,\dots,\abs{\mathcal{I}^c_{m}}\} \to \mathcal{I}^c_{m}$ be the bijection such that 
	\[
	\big| z_{\smash{ \pi_{\zeta}(1) }} - \zeta \big| \leq \dots \leq \big| z_{\smash{\pi_{\zeta} ( | \mathcal{I}^c_{m} | )}} - \zeta \big|,
	\]
	then $\mathcal{A}^{m,k,n}(\zeta,z^n)\coloneqq \{\pi_{\zeta}(1), \ldots, \pi_{\zeta}(k)\}$.
	Hereafter we use $\mathcal{A}^m(\zeta)$ instead as the remaining parameters can be understood from the context.
	In particular, we note that $\mathcal{A}^0(\zeta)$ implies that the neighbors are chosen from all points $z^n$ since $\mathcal{I}_0=\emptyset$ and $\mathcal{I}^c_{0} = \{1,\dots,n\}$. 
\end{definition}

According to the previous definition, the product batch with $b'= m k$ samples is defined as
\begin{align}
\mathcal{B}_\textnormal{prod}^{b'} \coloneqq \big\{ (X_{j(i)},Y_i,Z_i) \mid i\in \mathcal{I}_m,\, j(i)\in \mathcal{A}^m(Z_i) \big\}.
\label{eq:prodBatch}
\end{align}
We refer to this sampling technique as \emph{isolated $k$-NN} in the sequel.

Similar to the $k$ nearest neighbors method, the complexity of \emph{isolated $k$-NN} relies on the particular implementation.
With a brute force approach, the time complexity of computing distances from each sample in $\mathcal{I}_{m}$ to $\mathcal{I}_{m}^c$ is $\mathcal{O}(dn)$ when $m\ll n$, and a trivial search among these distances to find the $k$ nearest neighbors costs an additional $\mathcal{O}(kn)$.
Thus, the total time and storage complexities for all $m$ samples in the isolated set $\mathcal{I}_{m}$ are $\mathcal{O}((d+k)mn)$ and $\mathcal{O}(dn)$, respectively.
However, if the dimension of the data is small ($d\ll\log n$) and the number of queries (in our case $m$) is large, we can use alternative methods such as k-d trees where pre-processing enables us to find neighbors more efficiently.
Using k-d trees data structure with a nearly balanced tree, the time complexity becomes $\mathcal{O}((dn+mk)\log n)$ and the storage, $\mathcal{O}(dn)$.

\subsection{Approximating \texorpdfstring{$f_\DV^*$, $f_\NWJ^*$, and $f_\LDR^*$}{the optimal functions}}

To estimate the optimal functions, it suffices to obtain the likelihood ratio $\frac{p(x,y,z)}{p(x|z)p(y,z)}$.
As suggested in~\cite{mukherjee2019ccmi_conf, molavipour2020conditional}, we use a feed-forward neural network to classify inputs from the joint and product batches. 
In this network, the input is a triple $(x,y,z)$ and the last layer is concatenated with a sigmoid function.
Let the network be parameterized with $\theta$, then the output of the neural network is denoted as $\omega_{\theta}(x,y,z)$, see Fig.~\ref{fig:net}.
As it will be clear later, in order to avoid unbounded values in the ratio of densities, the output of the sigmoid function is clipped\footnote{It has been observed that such clipping also controls the bias--variance trade-off of the estimator~\cite{song2019understanding}. In our case, choosing $\tau$ closer to zero decreases the bias and allows for the estimation of large values of CMI while it also increases the variance of the estimation.} to the interval $[\tau,1-\tau]$ for $0<\tau< \frac{1}{2}$.

The binary cross-entropy loss is chosen as the objective function to optimize the network.
Define $q(x,y,z)$ (or simply $q$) as the batch type associated to an input, where $q=1$ and $q=0$ represent the joint and product batch, respectively.
Then we have the following definition for the cross-entropy loss.

\begin{definition}
	Given a function $\omega: \mathcal{X}^3 \to [0,1]$, the \emph{expected cross-entropy loss} is defined as:
	\begin{multline}
	L(\omega) \coloneqq -\ex_{p(q)p(x,y,z|q)} \big[ Q\log \omega(X,Y,Z)\\
	+ (1-Q)\log(1-\omega(X,Y,Z)) \big].
	\end{multline}
\end{definition}

The pointwise minimizer of $L(\omega)$ can be used to compute the desired likelihood ratio and, accordingly, the optimal functions, as we see next. Note that unlike $\omega_\theta$, the function $\omega$ is not restricted by any parameterization.

\begin{lemma}\label{lemma:optim_w}
	Let $\omega^*$ be the minimizer of the expected cross-entropy loss $L(\omega)$ and let $p(q=1)=p_1$, then
	\begin{align}
	\Gamma^*(x,y,z) \coloneqq \frac{1-p_1}{p_1}\frac{\omega^*(x,y,z)}{1-\omega^*(x,y,z)} =\frac{p(x,y,z)}{p(x|z)p(y,z)}.
	\label{eq:optim_w}
	\end{align}
\end{lemma}

\input{net_fig}

Using Lemma~\ref{lemma:optim_w}, the optimal functions~\eqref{eq:optimal_f_DV}, \eqref{eq:optimal_f_NWJ}, and~\eqref{eq:optim_f_LDR} can be evaluated as below,
\begin{align}
f^*_{\DV}(x,y,z) &\coloneqq C+\log \Gamma^*(x,y,z)\nonumber\\
f^*_{\NWJ}(x,y,z) &\coloneqq 1+\log \Gamma^*(x,y,z)\nonumber\\
f^*_{\LDR}(x,y,z) &\coloneqq \log \Gamma^*(x,y,z).\label{eq:optimal_f_2}
\end{align} 
However, there are restrictions to obtain \eqref{eq:optimal_f_2}. 
First, the optimization to achieve $\omega^*$ is performed on $L(\omega_\theta)$ over the parameterized networks $\omega_\theta$, as searching over all functions is infeasible. Second, since the densities are not available, the expectations in $L(\omega_\theta)$ are approximated with sample averages.

\begin{definition}
	Consider a neural-based classifier to be trained with sample batches $\mathcal{B}_\textnormal{joint}^{b}$ and $\mathcal{B}_\textnormal{prod}^{b'}$ such that 
	\begin{equation*}
	p_1=\frac{b}{b+b'},
	\end{equation*}
	then the \emph{empirical cross-entropy loss} is defined as:
	\begin{align}
	L_{\emp}(\omega_\theta) \coloneqq p_1 L_b^1(\omega_\theta)+ (1-p_1) L_{b'}^2(\omega_\theta),\label{eq:L2b_def}
	\end{align}
	where
	\begin{align}
	L_b^1(\omega_\theta) &\coloneqq -\frac{1}{b}\sum\limits_{(x,y,z)\in \mathcal{B}_\textnormal{joint}^{b}}  \log \omega_\theta(x,y,z)\nonumber\\
	L_{b'}^2(\omega_\theta) &\coloneqq -\frac{1}{b'}\sum\limits_{(x,y,z)\in \mathcal{B}_\textnormal{prod}^{b'}} \log\big( 1-\omega_\theta(x,y,z) \big).\label{eq:Lb12}
	\end{align} 
\end{definition}

Let $\hat\theta$ be the minimizer of $L_{\emp}(\omega_\theta)$, according to the previous definition, and define
\begin{align}
\hat\Gamma(x,y,z) \coloneqq \frac{1-p_1}{p_1} \frac{\omega_{\hat\theta}(x,y,z)}{1-\omega_{\hat\theta}(x,y,z)}.
\label{eq:Gamma_hat}
\end{align}
With a sufficiently large number of samples, $n$, and a proper tuning of the hyper-parameters of the network, $\hat\Gamma$ is close to $\Gamma^*$ with high probability and the variational bounds for CMI can be estimated as:
\begin{align}
\hat I_{\DV}^{n,\hat\theta} &\coloneqq \frac{1}{b}\sum\nolimits_{(x,y,z)\in\mathcal{B}_\textnormal{joint}^b} \log \hat\Gamma(x,y,z) \nonumber\\
&\quad -\log \frac{1}{b'} \sum\nolimits_{(x,y,z)\in\mathcal{B}_\textnormal{prod}^{b'}} \hat\Gamma(x,y,z),\nonumber\\ 
\hat I_\NWJ^{n,\hat\theta} &\coloneqq 1+ \frac{1}{b}\sum\nolimits_{(x,y,z)\in\mathcal{B}_\textnormal{joint}^b} \log \hat\Gamma(x,y,z)\nonumber\\
&\quad -\frac{1}{b'} \sum\nolimits_{(x,y,z)\in\mathcal{B}_\textnormal{prod}^{b'}} \hat\Gamma(x,y,z).\label{eq:est_CMI}
\end{align} 
Similarly, the estimation based on LDR can be obtained as:
\begin{align}
\hat I_\LDR^{n,\hat\theta}\coloneqq \frac{1}{b}\sum\nolimits_{(x,y,z)\in\mathcal{B}_\textnormal{joint}^b} \log \hat\Gamma(x,y,z). \label{eq:est_CMI_LDR}
\end{align}

There are two important caveats in computing these estimators.
First, in training the classifier, it is desired to have $p_1=\frac{1}{2}$ to avoid overfitting towards one of the classes. However, the prior can become biased due to the different resampling of the joint and product batches.
Second, to implement the cross-validation, the final estimation is averaged over $T$ trials where the train and test batches are re-sampled each time.
This has been advocated in~\cite{belghazi2018mine, mukherjee2019ccmi_conf} to control the variance of the estimation.
Note that with the averaging over multiple trials, the final DV estimator is no longer a lower bound for the CMI~\cite{molavipour2020conditional}. 
The steps of our proposed method are stated in Algorithm~\ref{alg:est}.
The functions jointBatch and isolated\_kNN in the algorithm execute~\eqref{eq:jointBatch} and~\eqref{eq:prodBatch}, respectively.

\begin{algorithm}[t]
	\SetAlgoLined
	\KwIn{$Data=\{(x_i,y_i,z_i)\}_{i=1}^n$, $T$, $b$, $b'$, $k$}
	Split $Data$ into $Train\_set$ and $Test\_set$\\	
	\For{t=1,\dots,T}
	{
		$\mathcal{B}^b_\textnormal{joint,train}\gets$ jointBatch($\,Train\_set\,,b$)\\\vspace{.1cm}
		$\mathcal{B}^{b'}_\textnormal{prod,train}\gets$ isolated\_kNN($\,Train\_set\,,b'\,,k$)\\\vspace{.1cm}
		$\omega_{\hat{\theta}}\gets$ Train the classifier with $\mathcal{B}^b_\textnormal{joint,train}, \mathcal{B}^{b'}_\textnormal{prod,train}$\\\vspace{.1cm}
		$\mathcal{B}^b_\textnormal{joint,test}\gets$ jointBatch($\,Test\_set\,,b$)\\\vspace{.1cm}
		$\mathcal{B}^{b'}_\textnormal{prod,test}\gets$ isolated\_kNN($\,Test\_set\,,b'\,,k$)\\\vspace{.1cm}		
		Compute $\hat I_\DV^{n,\hat\theta,t}$, $\hat I_\NWJ^{n,\hat\theta,t}$, and $\hat I_\LDR^{n,\hat\theta,t}$ using $\omega_{\hat{\theta}}$, $\mathcal{B}^b_\textnormal{joint,test}$, and $\mathcal{B}^{b'}_\textnormal{prod,test}$ as in \eqref{eq:est_CMI} and \eqref{eq:est_CMI_LDR}
	}
	$\hat I_\mathtt{est}^{n,\hat\theta}\gets \frac{1}{T}\sum_{t=1}^{T}  I_\mathtt{est}^{n,\hat\theta,t}$ for $\mathtt{est}=$'DV', 'NWJ', 'LDR'  \\ \vspace{.1cm} 
	\Return $\hat I_\DV^{n,\hat\theta}$, $\hat I_\NWJ^{n,\hat\theta}$ , $\hat I_\LDR^{n,\hat\theta}$
	\caption{Estimation of $I(X;Y|Z)$}
	\label{alg:est}
\end{algorithm}

\section{Main Results}\label{sec:main}

In this section, we discuss the consistency of the estimators $\hat I_\DV^{n,\hat\theta}$, $\hat I_\NWJ^{n,\hat\theta}$, and $\hat I_\LDR^{n,\hat\theta}$, which in general relies on two things:
\begin{itemize}
	\item The empirical sums in \eqref{eq:Lb12}, \eqref{eq:est_CMI}, and \eqref{eq:est_CMI_LDR} are concentrated around their expected values.
	For instance, this implies that for any $\theta$, $L_{\emp}(\omega_\theta)$ falls in the neighborhood of $L(\omega_\theta)$ with high probability, if certain conditions hold.
	
	\item The hyper parameters of the feed-forward neural network can be found such that with a perfect optimizer over parameters $\theta$, one can desirably approximate $\omega^*$, and accordingly $f^*_\DV$, $f^*_\NWJ$, and $f^*_\LDR$.   
\end{itemize} 
In the following, we first show that the empirical average (using \textit{isolated} $k$-NN re-sampled data) of any function $g$ with bounded codomain converges to the expectation with respect to $p(x|z)p(y,z)$.
Then, for our neural estimators of CMI, we exploit this result and the universal functional approximation theorem~\cite{hornik1989multilayer} to show that our estimators are consistent.

\subsection{Concentration results}

To obtain a high confidence concentration bound, let us make the following assumption on the value of $k$.

\begin{assumption}\label{Assum:knn}
	Consider $k(n)=\varTheta(n^{\frac{1}{2}+\epsilon_0})$ for some $\epsilon_0>0$ and $n-k(n)\geq m(n)\geq k(n)$. Then we
	select $b'(n)=m(n)k(n)$ samples to create the product batch\footnote{It is worth noting that $b'$ is then upper bounded by $m(n-m)$, as by choosing $m$ indices for $\mathcal{I}_m$, we are left with at most $n-m$ samples from which to choose the neighbors, i.e., $k\leq n-m$.} in the \emph{isolated} $k$-NN method, using $k(n)$ neighbors and isolation set of size $m(n)$, as described in Definition~\ref{def:isoKnn}. 
	On the other hand, for the joint batch, assume $b(n)=\varTheta(n)$.
	Hereafter we continue to use the notation $b$, $b'$, $m$, and $k$, except where the dependency with $n$ is important. 	
\end{assumption}

\begin{remark}
	Assumption~\ref{Assum:knn} is required to prove convergence.
	If we choose $m(n)=k(n)=n^{\frac{1}{2}+\epsilon_0}$, the size of the product batch becomes larger than $n$, i.e., $b'(n)=n^{1+2\epsilon_0}$.
	This is not an issue because the \emph{isolated} $k$-NN technique enables us to construct batches of size larger than $n$, since the technique re-samples and mixes the original data, thus creating new samples.
	Nevertheless, we will show in the experimental results that even a smaller choice of $k$ can yield a good estimation performance (e.g., see Fig.~\ref{fig:iso-diff}, where $n=8\mathrm{e}4$ while $k=2$). 
	Additionally, to balance the size of the joint and product batch, we can adjust the number of samples used to create the product batch. 
	So in the example above, if we only use $\tilde n=n^{1/(1+2\epsilon_0)}$ samples and choose 	$m(\tilde n)=k(\tilde n)=\tilde{n}^{\frac{1}{2}+\epsilon_0}$, then $b'(n)=n$. 
\end{remark}

Now to address the concentration of the empirical average over samples taken with the \textit{isolated $k$-NN} technique, we introduce the following theorem. 

\begin{theorem}\label{th:convergence}
	Let $g(x,y,z):\mathcal{X}^3 \to\mathbb{R}$ be any function such that $g^{min} \leq g(x,y,z)\leq g^{max}$, and $M\coloneqq\max\left\{\abs{g^{min}},\abs{g^{max}}\right\}$ is finite.
	Consider 
	\begin{align}
	\hat g(x^n,y^n,z^n)\coloneqq\frac{1}{m}\sum_{i\in\mathcal{I}_m}\frac{1}{k} \sum_{j\in\mathcal{A}^m(z_i)} g(x_j,y_i,z_i),\label{eq:g_hat_def}
	\end{align}
	with $k$ and the set $\mathcal{A}^m$ as defined in Assumption~\ref{Assum:knn} and Definition~\ref{def:isoKnn}, respectively.
	Then, for any $\epsilon>0$ there exists an integer $n_0$ such that for $n>n_0$ and $m\leq n$,
	\begin{align}
	\prB{\abs{\hat g(x^n,y^n,z^n) - \ex_{\dprod}[g(X,Y,Z)]}\geq 3\epsilon}\nonumber\\
	\leq \mdel{1}(\epsilon,c,M),
	\end{align}
	where $\mdel{1}$ is defined in Table~\ref{tab:param}, $c \coloneqq g^{max}-g^{min},$
	and $\gamma_d$ is the minimal number of cones centered at the origin, of angle $\pi/6$, that cover $\mathbb{R}^d$.
	
	\begin{proof}
		See Appendix~\ref{App:proof_th_convergence}.		
	\end{proof}
\end{theorem}

\begin{remark}
	Note that $\lim_{n\to\infty} k(n)^2/n=\infty$, according to Assumption~\ref{Assum:knn}.
	Additionally, $n-m(n)=\varTheta(n)$, which concludes that $\lim_{n\to\infty} \mdel{1}(\epsilon,c,M)= 0$.
\end{remark}

Theorem~\ref{th:convergence}, in conjunction with Hoeffding's inequality, leads to the concentration bound on $L_{\emp}(\omega_\theta)$ found in the following proposition.
This result is crucial in order to later show that $\omega_{\hat{\theta}}$ is close to $\omega^*$, where we recall that $\hat{\theta}$ is the minimizer of the empirical loss.

\begin{proposition}\label{Prop:convergence_L}
	Let Assumption~\ref{Assum:knn} hold.
	Then, for any $\mu>0$ and any $\theta$ there exists $n_0$ such that for $n>n_0$,
	\begin{align}
	\prB{ \Big| L_{\emp}(\omega_\theta)-L(\omega_\theta) \Big| \geq \mu}\leq \mdel{3}(\mu),
	\end{align}
	where $\mdel{3}$ is defined in Table~\ref{tab:param}. 
	
	\begin{proof}
		See Appendix~\ref{App:proof_prop_L}.
	\end{proof}
\end{proposition}

\subsection{Consistency of the estimators}

To study the consistency of $\hat I_\DV^{n,\hat{\theta}}$, $\hat I_\NWJ^{n,\hat{\theta}}$, and $\hat I_\LDR^{n,\hat{\theta}}$ in estimating $I(X;Y|Z)$, we make some further assumptions. 

\begin{assumption}\label{Assum:prob}
	There exist $0<\alpha<\beta<\infty$ such that for any finite input $x$, $y$, $z$ $\in\mathcal{X}^3$, the values of $\djoint$ and $\dprod$ are both constrained to the interval $[\alpha,\beta]$.
\end{assumption}

\begin{remark}
	The constraint stated in Assumption~\ref{Assum:prob} helps us show the consistency of our proposed estimators.
	However, most of our experiments are with Gaussian data, which does not fulfill the requirements of this assumption.
	Nonetheless, given the good simulation results, we believe that another less restrictive assumption on the densities could replace Assumption~\ref{Assum:prob}.
\end{remark}

\begin{assumption}\label{Assum:NN_Lip}
	The classifier is parameterized with $\theta\in\Theta$ where $\Theta\subset\mathbb{R}^h$ and $h$ is the number of parameters in the neural network.
	Also $\| \theta\|_2\leq K$ for a constant $K$ and the output of the classifier is $B$-Lipschitz with respect to $\theta$.
	The hyper-parameters $h$, $K<\infty$ depend on the approximation power of the neural network to classify the dataset, and they can be determined according to the complexity of the dataset, the structure of the neural network, and others. 
\end{assumption}

	\begin{remark}
		By restricting the output of the classifier to be Lipschitz continuous with respect to $\theta$, the activation functions of the neural network must be differentiable (e.g., softplus).
		Nonetheless, we use rectified linear units (ReLU) in our experiments, similar to \cite{belghazi2018mine, poole2018variational, mukherjee2019ccmi_conf}, and obtain a desirable estimation performance.
		Note that the softplus function, defined as $f(x)=\frac{1}{t} \ln(1+e^{tx}),$ is equivalent to ReLU, asymptotically as $t\to\infty$.
		However, the use of the ReLU function is encouraged over the softplus~\cite{goodfellow2016deep}.  
	\end{remark}

While Assumption~\ref{Assum:knn} guarantees that $\mdel{1}(\epsilon,c,M), \mdel{2}(\epsilon), \dots,\allowbreak \delta_7(\epsilon)$ tend to zero asymptotically as $n\to\infty$, in order to obtain a concentration bound, the sample size $n$ needs to be larger than a certain threshold.
This value is determined by the true density $p(x,y,z)$ and the hyper-parameters of our setup, and it is stated in the following assumption.\footnote{The sample complexity for the neural estimator of MI has been discussed in~\cite[Theorem~3]{belghazi2018mine} and recently revisited in the fourth online version of~\cite{mcallester2018formal}.
Similar results exist for the classifier estimator for the CMI in~\cite[Lemma~5]{mukherjee2019ccmi_conf}.}

\begin{assumption}\label{Assum:b}
	For given $\epsilon^*>0$ and $\delta^*>0$, we assume that $n$ is large enough such that the following conditions hold: 
	
	\begin{align*}
	\mdel{4}\left(\frac{\epsilon}{8}\right) +\delta_i^n(\epsilon^*) \leq\delta^*,\ \textnormal{ for }\ i\in\{5,6,7\},
	\end{align*}
	where $\epsilon$ and $\mdel{i}(\cdot)$ are defined in Table~\ref{tab:param}.
	Note that by the continuity of the functions $\mdel{i}(\cdot)$ and their asymptotic behavior, finding such an $n$ is feasible.
\end{assumption}

Now we are able to express the consistency of our estimators in terms of concentration bounds in the following theorem.

\begin{theorem}\label{th:consistency}
	Let Assumptions~\ref{Assum:knn}, \ref{Assum:prob}, \ref{Assum:NN_Lip}, and~\ref{Assum:b} hold and $0<\tau<\min\{\frac{1}{2},p_1\}$.
	Then, given $\epsilon^*, \delta^*>0$, there exists an integer $n^*$ such that for all $n>n^*$,
	\begin{align}
	\prB{ \Big| \hat I_{\mathtt{est}}^{n,\hat\theta} - I(X;Y|Z) \Big| \geq \epsilon^*} &\leq \delta^*,
	\end{align}
	where `$\mathtt{est}$' can be replaced with `$\DV$', `$\NWJ$', or `$\LDR$'.
	\begin{proof}
		See Appendix~\ref{appendix:Proof_Th2}.
	\end{proof}
\end{theorem}

\begin{remark}
	Note that the proper choice of the hyper-parameters of the network and the value of $n^*$ crucially relies on the true underlying density, and thus the bounds are not universal in that sense.
	This has been emphasized in~\cite[Remark~7]{pichler2020estimation} where the authors discuss required precautions for the neural estimator in~\cite{belghazi2018mine}.
\end{remark}

\begin{table}\centering
	\caption{Table of parameters.}
	\label{tab:param}
	
	\begin{tabular}{l}\hline\\[-.25cm]
		$\mdel{1} (\epsilon,c,M) \coloneqq 2\exp\! \big( \sm{\frac{-2\epsilon^2 k^2}{nc^2}} \big) + 2\exp\! \big( \sm{\frac{-2\epsilon^2 k^2}{(n-m)c^2}} \big) + \exp\! \big( \frac{-(n-m)\epsilon^2}{8 M^2\gamma_d^2} \big)$\\ [.2cm]
		$\mdel{2}(\epsilon)\coloneqq \mdel{1}\big( \epsilon, \sm{\log\frac{1-\tau}{\tau}}, -\log\tau \big)$\\ [.1cm]
		$\mdel{3}(\epsilon)\coloneqq\mdel{2}\big(\frac{\epsilon}{3-2p_1}\big) + 2\exp\bigg(\frac{ -2\,  b(n)\,\epsilon^2}{\big((3-2 p_1)\log\frac{1-\tau}{\tau}\big)^2}\bigg)$ \\ [.1cm]
		$\mdel{4}(\epsilon)\coloneqq\Big(\frac{4BK\sqrt{h}}{\tau\epsilon}\Big)^h\,\mdel{3}(\epsilon)$ \\ [.2cm]
		$\mdel{5}(\epsilon)\coloneqq\mdel{1}\big(\frac{(1-p_1)\epsilon\,\tau}{2\tau+6p_1-8p_1\tau},\sm{\frac{1-p_1}{p_1}\frac{1-2\tau}{\tau(1-\tau)}},\frac{1-p_1}{p_1}\frac{1-\tau}{\tau}\big)$\\ [.2cm]
		$\hspace{1.05cm} {}+ 2\exp\bigg(\frac{- b(n)(1-p_1)^2 \epsilon^2\,\tau^2}{2\big((2\tau+6p_1-8p_1\tau)\log\frac{1-\tau}{\tau}\big)^2}\bigg)$\\ [.3cm]
		$\mdel{6}(\epsilon)\coloneqq\mdel{1}\big(\frac{\epsilon}{8},\sm{\frac{1-p_1}{p_1}\frac{1-2\tau}{\tau(1-\tau)}},\frac{1-p_1}{p_1}\frac{1-\tau}{\tau}\big)+2\exp\bigg(\frac{- b(n) \epsilon^2}{128\big(\log\frac{1-\tau}{\tau}\big)^2}\bigg)$ \\ [.2cm]
		$\mdel{7}(\epsilon)\coloneqq 2\exp\bigg(\frac{- b(n) \epsilon^2}{8\big(\log\frac{1-\tau}{\tau}\big)^2}\bigg)$ \\ [.2cm]
		$\eta\coloneqq\frac{\tau^3(1-\tau)\,\epsilon^*}{2(2\tau^2-2\tau+1)\beta}$\\ [.2cm]
		$\epsilon\coloneqq\big(\frac{\eta}{1-\tau}\big)^2 \frac{\alpha}{2 \lambda(\mathcal{X})}$\\ [.1cm]		
		\hline
	\end{tabular}
\end{table}

\section{Experiments}\label{sec:simul}

In this section, we compare our technique\footnote{Code: \url{https://github.com/smolavipour/CMI_Neural_Estimator}} with the state-of-the-art approach proposed in \cite{mukherjee2019ccmi_conf}, which we refer to as \textit{MI-Diff} (also known as the CCMI method) since the method is based on computing the CMI with the difference of two MI terms, $I(X;Y,Z)$ and $I(X;Z)$, as in~\eqref{eq:CMI_2_MI}.
Each MI term is then estimated by utilizing a neural network classifier with a similar structure as our method and training with the proper joint and product batches.
In contrast to the \textit{isolated $k$-NN} method, the construction of batches in \textit{MI-Diff} is straightforward. The joint batches are created similar to \eqref{eq:jointBatch}, while the product batches for $I(X;Y,Z)$ and $I(X;Z)$ are constructed by taking $b$ random indices separately from $x^n$ and $(y^n,z^n)$. 
In particular for $I(X;Z)$, the product batch in the \textit{MI-Diff} method is:
\begin{align}
\mathcal{B}_\textnormal{prod}^b \coloneqq \big\{(X_i,Z_j) \mid i\in \mathcal{I}^{(1)}_b, j\in \mathcal{I}^{(2)}_b \big\},
\label{eq:prodBatch_MIDiff}
\end{align}
where $\mathcal{I}^{(1)}_b$ and $\mathcal{I}^{(2)}_b$ are random indices selected separately in $\{1,\dots,n\}$. Similarly for $I(X;Y,Z)$ the product batch is:
\begin{align}
\mathcal{B}_\textnormal{prod}^b \coloneqq \big\{(X_i,Y_j,Z_j) \mid i\in \mathcal{I}^{(1)}_b, j\in \mathcal{I}^{(2)}_b \big\}.
\label{eq:prodBatch2_MIDiff}
\end{align}

Initially, we verify the approximation power and consistency of the estimators in two scenarios where the CMI is either non-zero (part~A) or zero (part~B). Additionally, we investigate the performance of our method as the dimension of data grows (part~C). The generative model that we use is defined as:
\begin{align}
X &\sim \mathcal{N}(0\,,\sigma^2_x\,\Sigma_d), \nonumber\\
Y &\sim \mathcal{N}(X,\sigma^2_y\,\Sigma_d), \nonumber\\
Z &\sim \mathcal{N}(Y,\sigma^2_z\,\Sigma_d). \label{eq:model}
\end{align}
In order to meet Assumption~\ref{Assum:prob}, we could use a truncated normal distribution by bounding the $\ell_2$ norm of the random variables.
However, slight deviations from this assumption do not significantly change the statistics of the generated dataset since the likelihood of observing a very large or low value is negligible.

\begin{figure}[t]
	\centering
	\includegraphics[width=.9\linewidth]{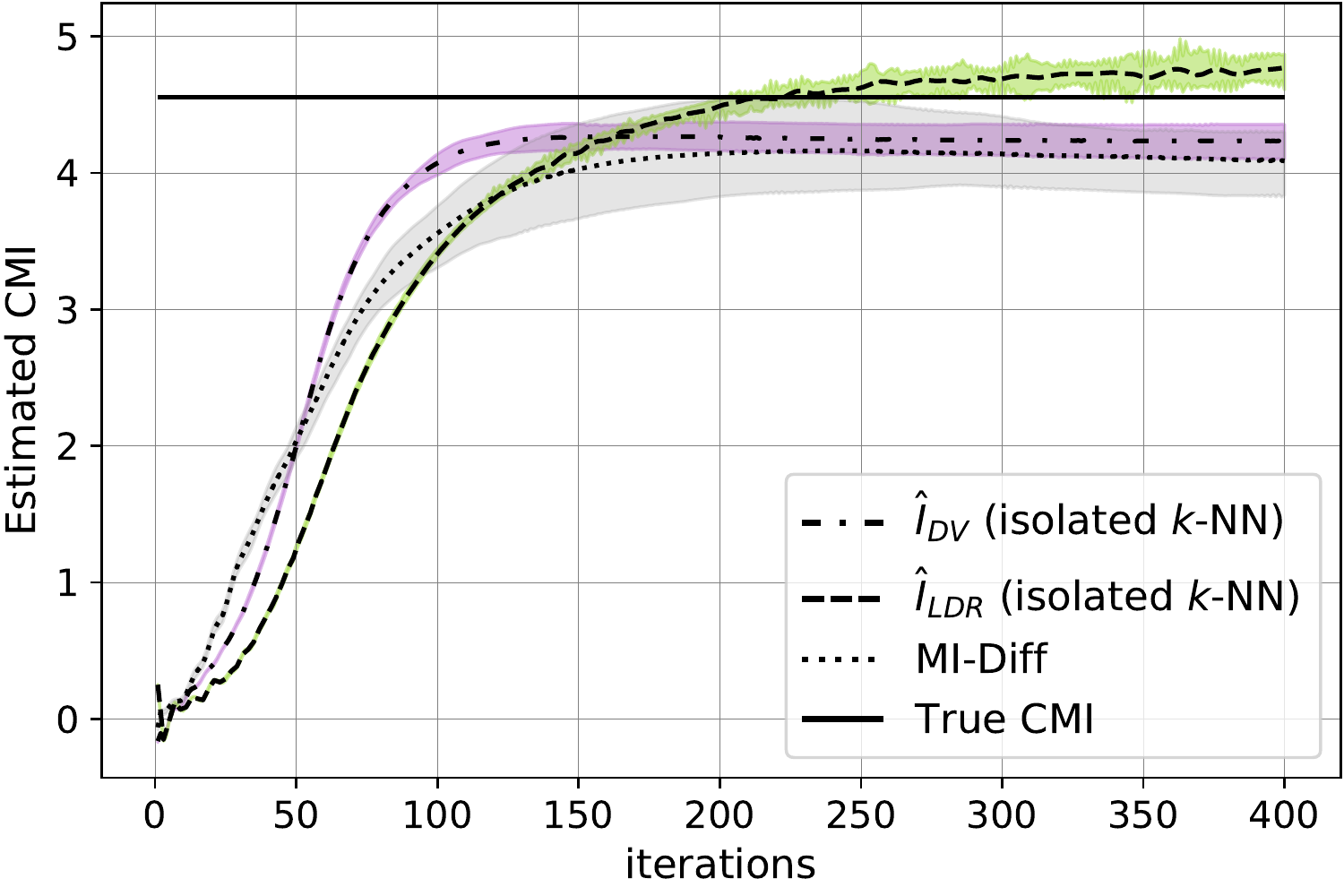}
	\caption{Evolution of the estimation of $I(X;Y|Z)$ over training iterations, using our estimators $\hat{I}^{n,\hat{\theta}}_{\DV}$ and $\hat I^{n,\hat{\theta}}_{\LDR}$  compared with the \textit{MI-Diff} method based on the DV bound. The input dataset contains $n=8\mathrm{e}{4}$ samples with $d=3$. 
	The shadows are based on the maximum and minimum values obtained at each iteration.}
	\label{fig:training}
\end{figure}

\subsection{Estimating \texorpdfstring{$I(X;Y|Z)$}{I(X;Y|Z)}}

Consider $\sigma_x=10$, $\sigma_y=1$, and $\sigma_z=5$ in our model~\eqref{eq:model}, and let $\Sigma_d=I_d$, i.e., the identity matrix of dimension $d$.
According to the model, we compute $I(X;Y|Z)$ as follows:
\begin{align*}
I(X;Y|Z) &= I(X;Y) -I(X;Z) \nonumber\\
&= \frac{d}{2} \log \bigg(1+ \frac{\sigma_x^2}{\sigma_y^2}\bigg) -\frac{d}{2} \log \bigg(1+ \frac{\sigma_x^2}{\sigma_y^2 + \sigma_z^2}\bigg).
\end{align*}

The evolution of the neural classifier's performance using a validation dataset during training is shown in Fig.~\ref{fig:training}, for $d=3$ and $k=2$.
We made a comparison between the estimators $\hat I^{n,\hat\theta}_\DV$ and $\hat I^{n,\hat\theta}_\LDR$ according to Algorithm~\ref{alg:est}, with the \textit{MI-Diff} method.
Both the MI-Diff and $\hat I^{n,\hat{\theta}}_{\DV}$ estimators converge after $E=200$ epochs, while $\hat I^{n,\hat{\theta}}_{\LDR}$ requires more iterations ($E\geq 300$) to converge.
Comparing the range of the estimations suggests that the DV and LDR estimators have lower variance compared to the \textit{MI-Diff}.

\begin{figure}[t]
	\centering
	\includegraphics[width=.9\linewidth]{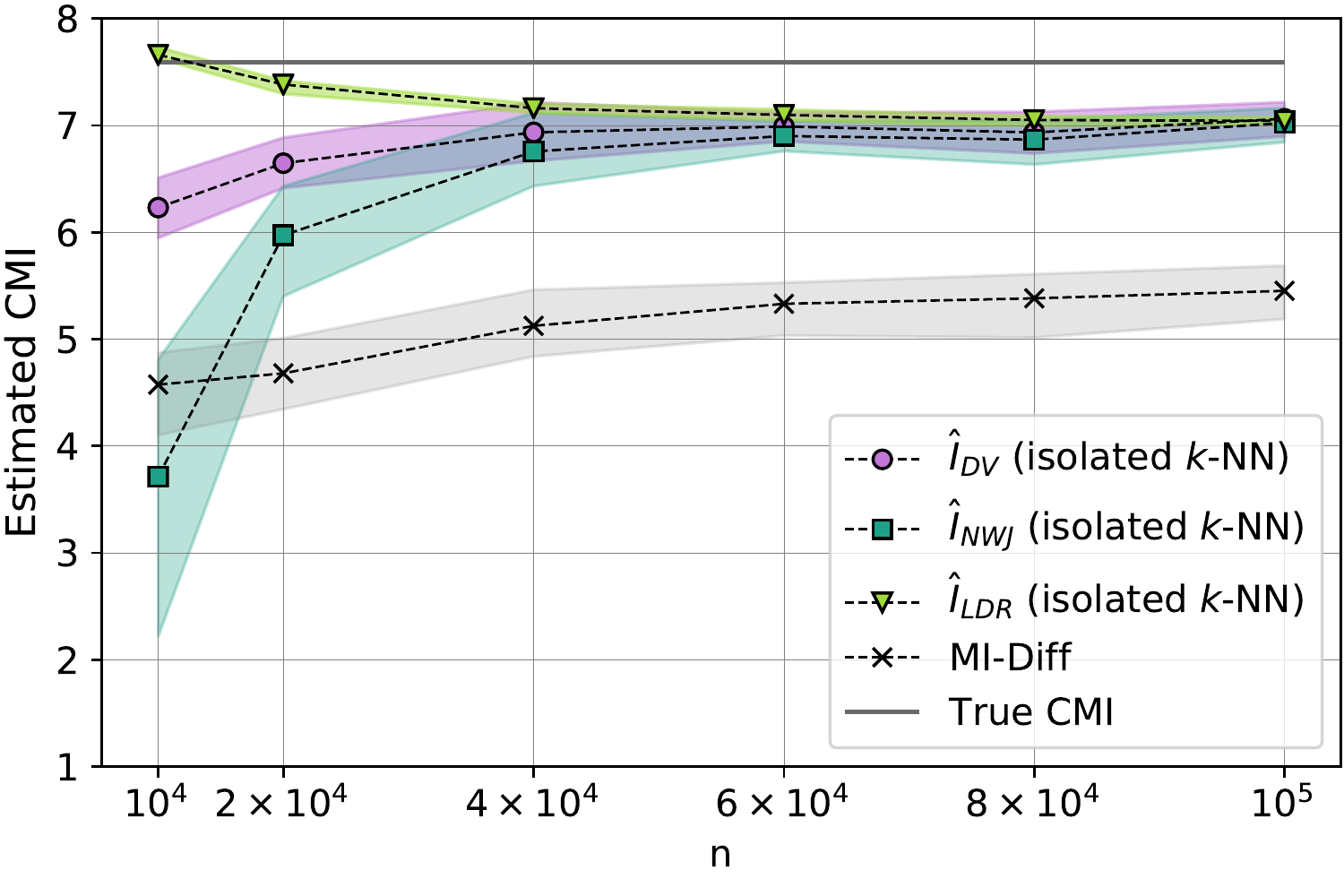}
	\caption{Comparison between our proposed estimators using \textit{isolated $k$-NN}, with $k=2$, and the \textit{MI-Diff} method to estimate $I(X;Y|Z)$, with $d=5$, $E=300$, and different values of $n$.}
	\label{fig:comp_d5}
\end{figure}

Next, we show significant improvements of our estimators compared with the \textit{MI-Diff} method when the dimension increases. 
In Fig.~\ref{fig:comp_d5}, a comparison of the estimators for CMI is depicted for $d=5$ in terms of sample size $n$. 
Our LDR estimator performs better than both DV and NWJ estimators in terms of bias and variance. Note that the LDR estimator is averaging the density ratio over samples in the joint batch as
$$ \frac{1}{b}\sum\nolimits_{(x,y,z)\in\mathcal{B}^b_\textnormal{joint}} \log \frac{p(x,y,z)}{p(x|z)p(y,z)}.$$%
For a small number of samples, typically the samples with high probability density $p(x,y,z)$ appear while by having more samples, the chance of observing odd samples with low probability increases, which compensates the total average.
This effect results in the LDR estimation to have a decreasing behavior by increasing $n$. On the other hand, the DV estimator is of the form
$$\frac{1}{b}\!\!\!\sum_{(x,y,z)\in\mathcal{B}^b_\textnormal{joint}} \!\!\! \log \frac{p(x,y,z)}{p(x|z)p(y,z)} - \log \frac{1}{b'}\!\!\! \sum_{(x,y,z)\in\mathcal{B}^{b'}_\textnormal{prod}} \frac{p(x,y,z)}{p(x|z)p(y,z)},$$%
where the second sum is dominated by the unlikely events.
So even by observing one odd event, the second term becomes very large. 
As $n$ increases, more typical samples are collected in the product batch and this effect disappears.

\begin{figure}[t]
	\centering
	\includegraphics[width=.9\linewidth]{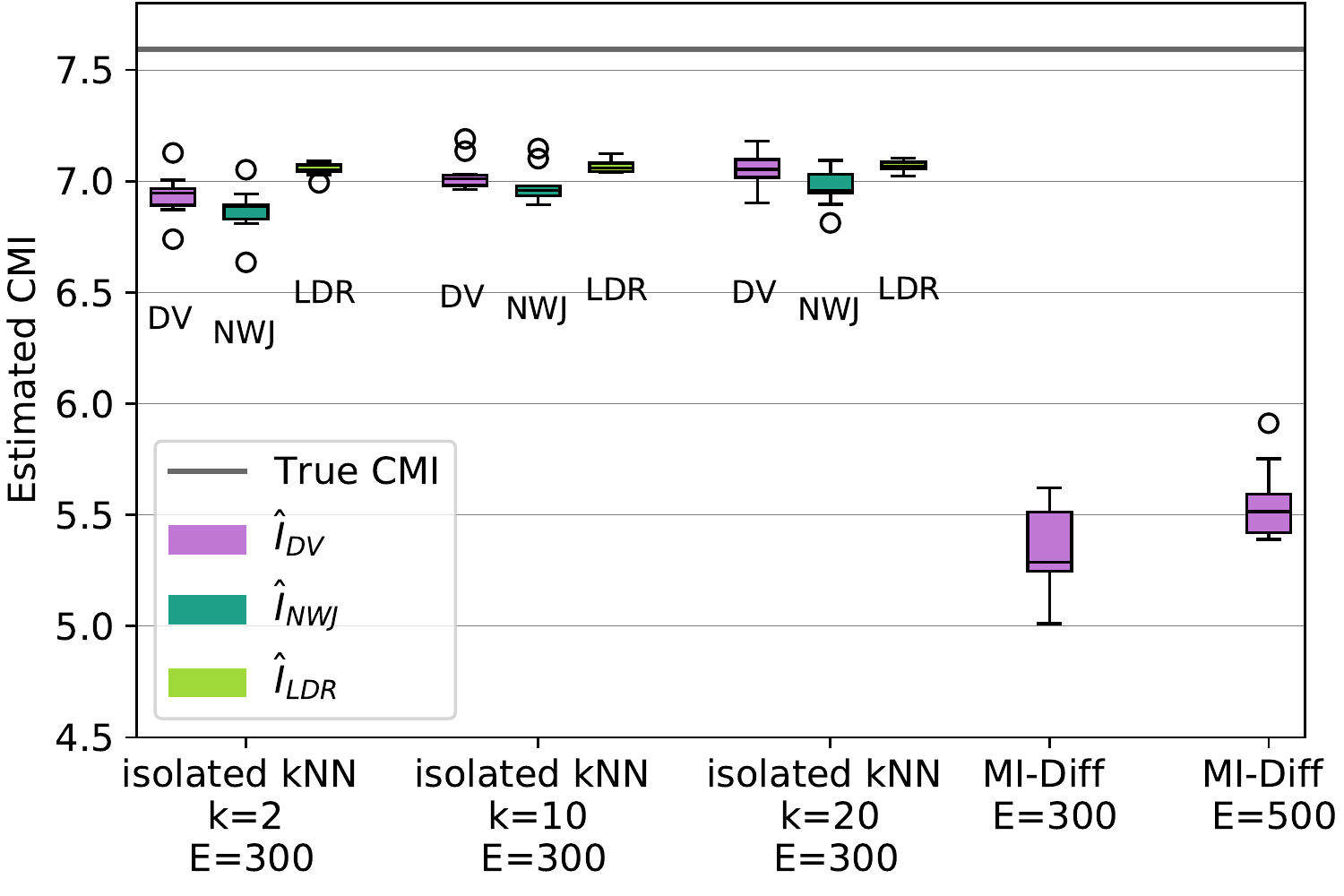}
	\caption{Comparison between the \textit{isolated $k$-NN} and the \textit{MI-Diff} methods to estimate $I(X;Y|Z)$, for an input dataset with $n=8\mathrm{e}{4}$ samples and $d=5$.}
	\label{fig:iso-diff}
\end{figure}

In Fig.~\ref{fig:iso-diff}, $I(X;Y|Z)$ is estimated for $n=8\mathrm{e}{4}$ and dimension $d=5$, with different choices of $k$.
Then the results are compared with the \textit{MI-Diff} method for estimating the DV bound.
It can be observed that sampling batches using \textit{isolated $k$-NN} improves the accuracy of the estimation.
To ensure that the training in the \textit{MI-Diff} method has been done with enough epochs, we repeated the experiment with more epochs as well.
Despite leveraging additional learning iterations, both accuracy and variance of our estimators are more desirable.

\begin{figure}[t]
	\centering
	\subfloat[DV estimation for different values of $k$ and $n$.]{
		\includegraphics[width=0.9\linewidth]{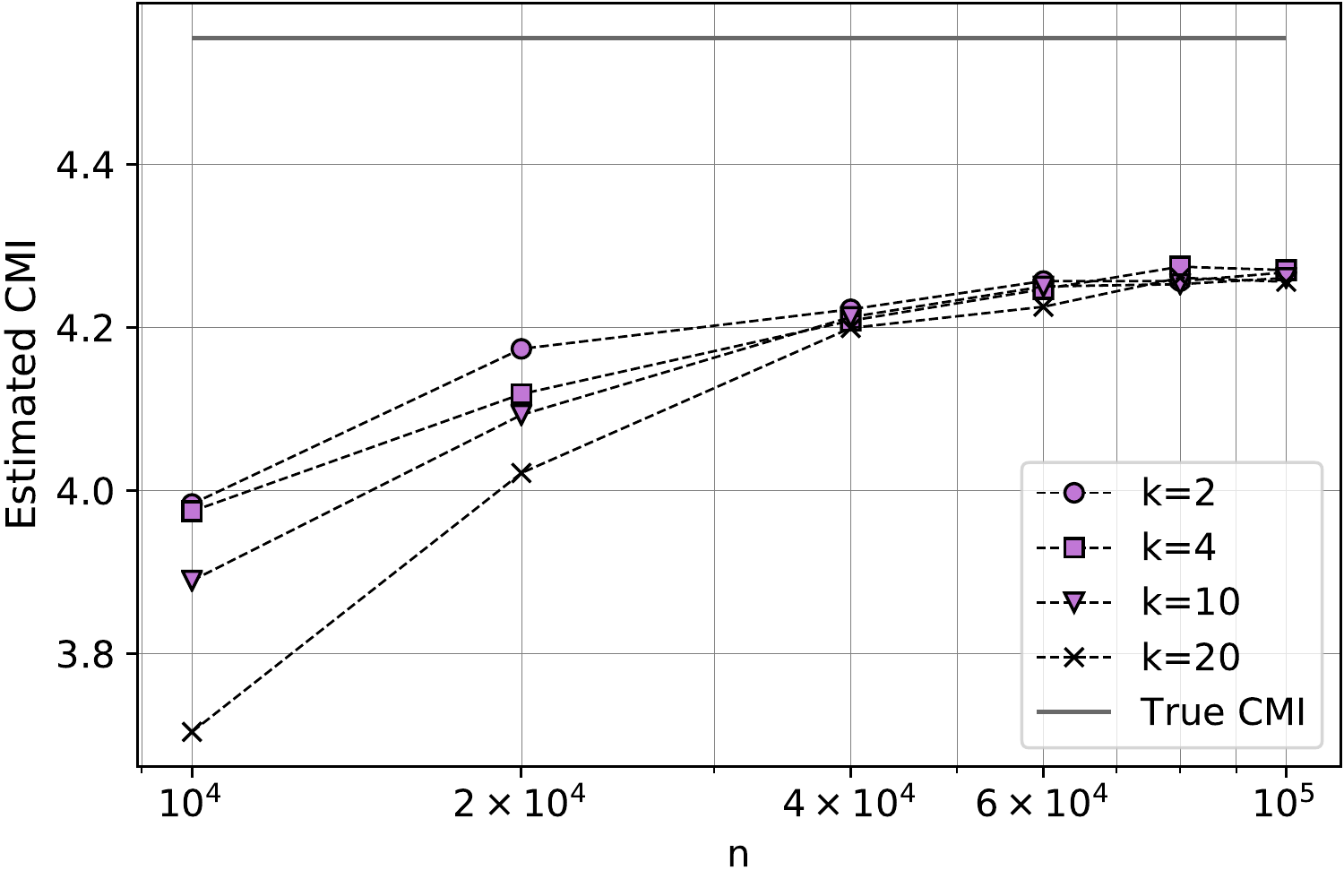}
		\label{fig:iso1}
	}\\
	\subfloat[DV and LDR estimations with a fixed $k/n$ ratio.]{
		\includegraphics[width=0.9\linewidth]{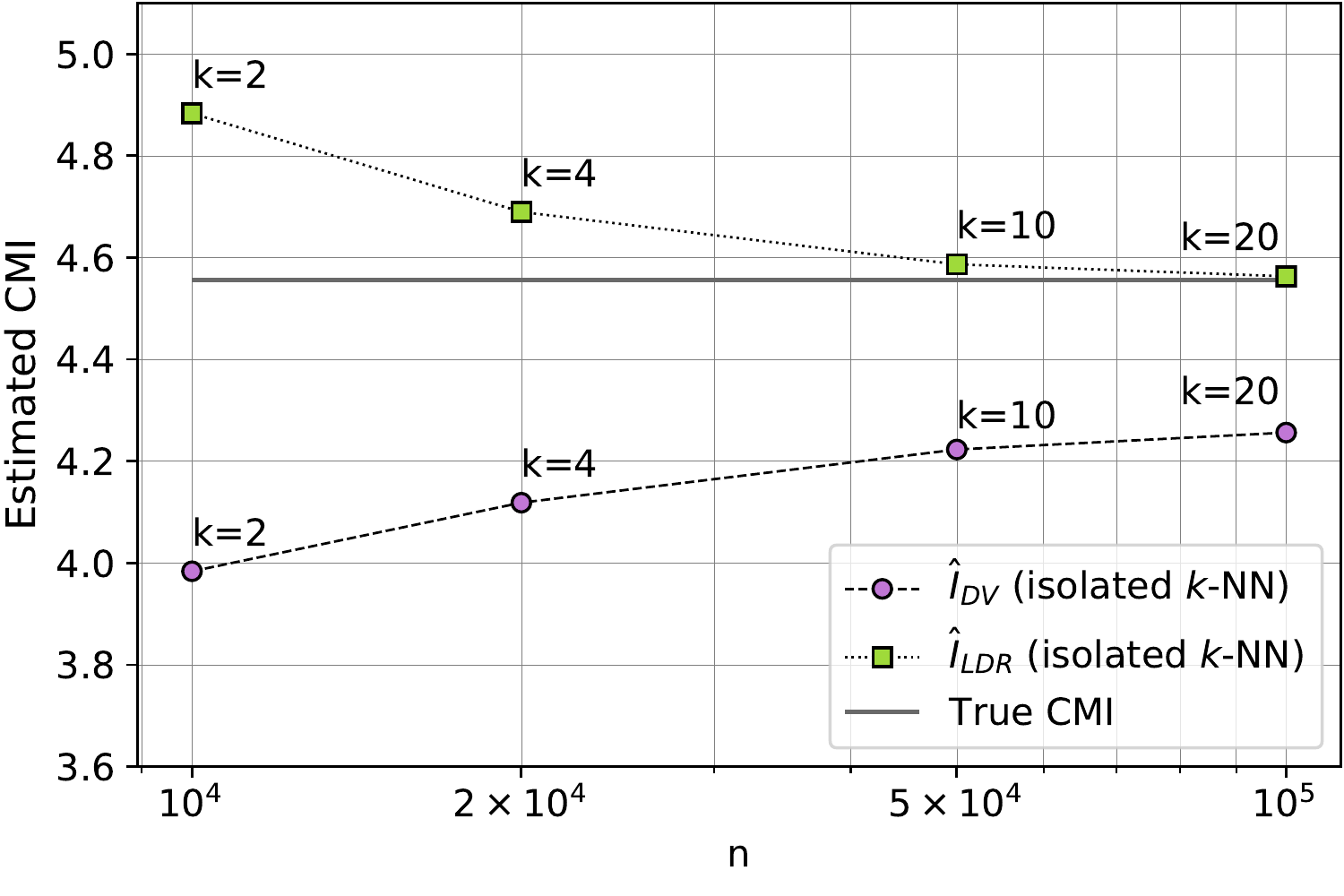}
		\label{fig:iso2}
	}	
	\caption{Estimated $I(X;Y|Z)$ using the \textit{isolated $k$-NN} technique, with $d=3$.}
	\label{fig:iso}
\end{figure}

As suggested in the \textit{isolated $k$-NN} method, increasing $k$ can improve the estimation if it is properly scaled with $n$ and $\lim\nolimits_{n\to\infty}k(n)/n=0$.
To investigate this, we compare the estimated CMI with $\hat I^{n,\hat{\theta}}_\DV$ for $d=3$ and different choices of $k$ and $n$ in Fig.~\ref{fig:iso1}.
In general, increasing the number of samples for a fixed $k$ results in a more accurate estimation, as shown in Fig.~\ref{fig:iso1}.
However, when the number of samples is fixed, choosing a larger $k$ worsens the estimation.
The reason is that with $n$ being fixed and $b=m\, k=\frac{n}{2}$, $m$ becomes smaller and there are less samples of $(y,z)$ to estimate the expectation $\ex_{p(y,z)}[\cdot]$ with sample average.
Nonetheless, this behavior can be resolved if $k=k(n)$ increases with $n$, and as a result $m$ can remain sufficiently large to obtain a desired accuracy.
This is illustrated in Fig.~\ref{fig:iso2} where with a fixed ratio of $k(n)/n$, the estimation improves by increasing $k$.

\begin{figure}[t]
	\centering
	\includegraphics[width=0.9\linewidth]{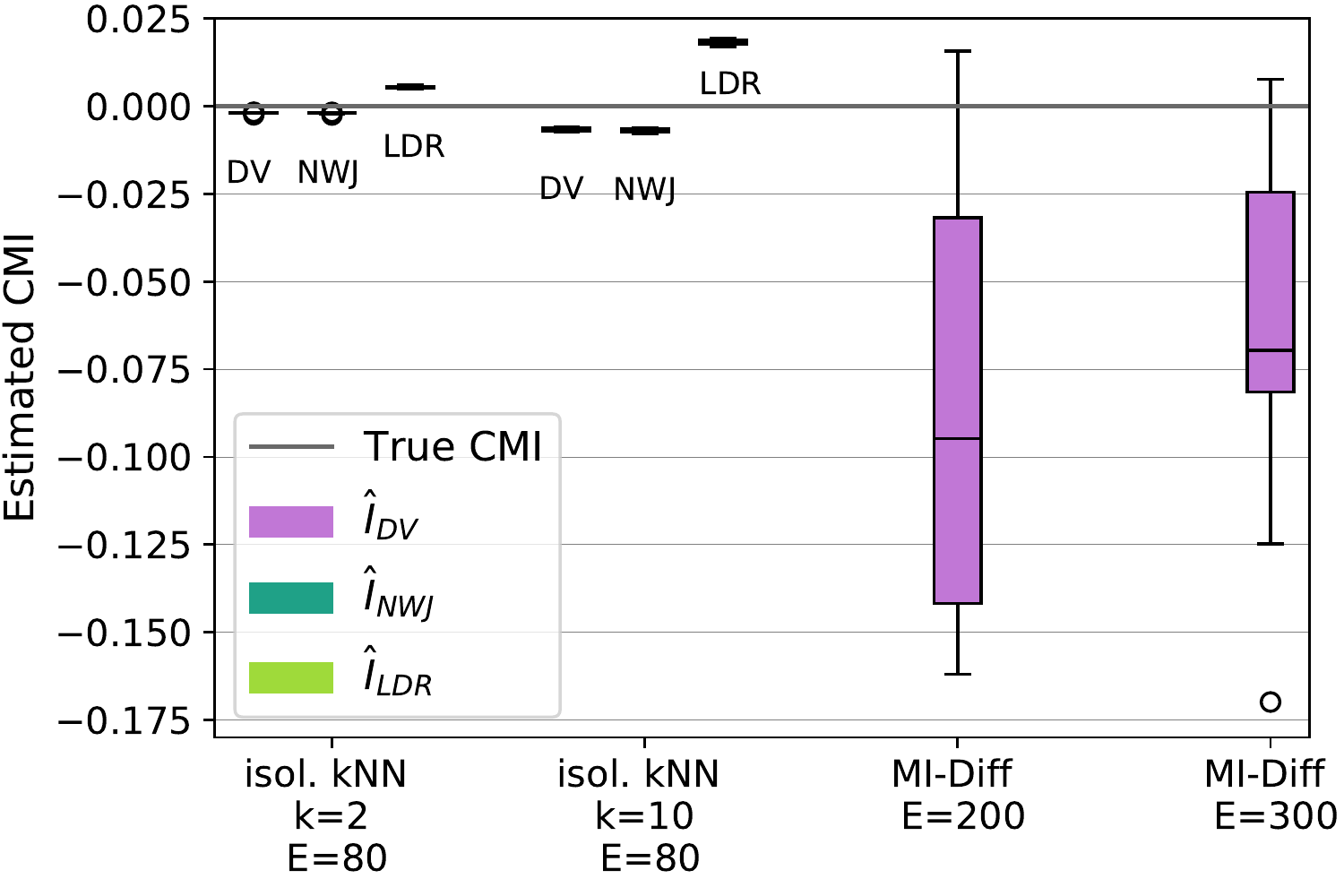}
	\caption{Comparison between the \textit{isolated $k$-NN} and the \textit{MI-Diff} methods to estimate $I(X;Z|Y)=0$, for $n=8\mathrm{e}{4}$ and $d=3$.}
	\label{fig:iso-diff_zero}
\end{figure}

\subsection{Estimating \texorpdfstring{$I(X;Z|Y)$}{I(X;Z|Y)} (zero CMI)}

A desirable estimator must be able to estimate both high and low values of CMI.
In this scenario, we test the ability of our estimator for zero CMI. 
Due to the Markov chain $X\to Y\to Z$ in the model \eqref{eq:model}, $I(X;Z|Y)=0$.
Consider the same choices for $\sigma_x$, $\sigma_y$,  $\sigma_z$, and $\Sigma_d$ as in previous part while $d=3$.
In Fig.~\ref{fig:iso-diff_zero} the box-plots are created by repeating Algorithm~\ref{alg:est} and \textit{MI-Diff} for 10 Monte Carlo trials and the sample size is $n=8\mathrm{e}{4}$. The results of \textit{isolated $k$-NN} are shown for $k=2$ and $10$. The performance of the \textit{MI-Diff} method is shown for different choices of number of epochs, while we exclude the case of $E=80$ due to poor estimation (median$<-0.3$).

It can be observed that our technique has the advantage of lower bias and variance compared with \textit{MI-Diff}. 
We note that, as explained earlier, the degrading performance by increasing $k$ is due to $n$ being fixed.
Furthermore, although the CMI is non-negative, we see that the estimation can become negative if the density ratio is not estimated properly.

\begin{figure}[t]
	\centering
	\subfloat[Estimation with $n=2\mathrm{e}{5}$ and $\tau=3\mathrm{e}{-5}$]{
		\includegraphics[width=0.9\linewidth]{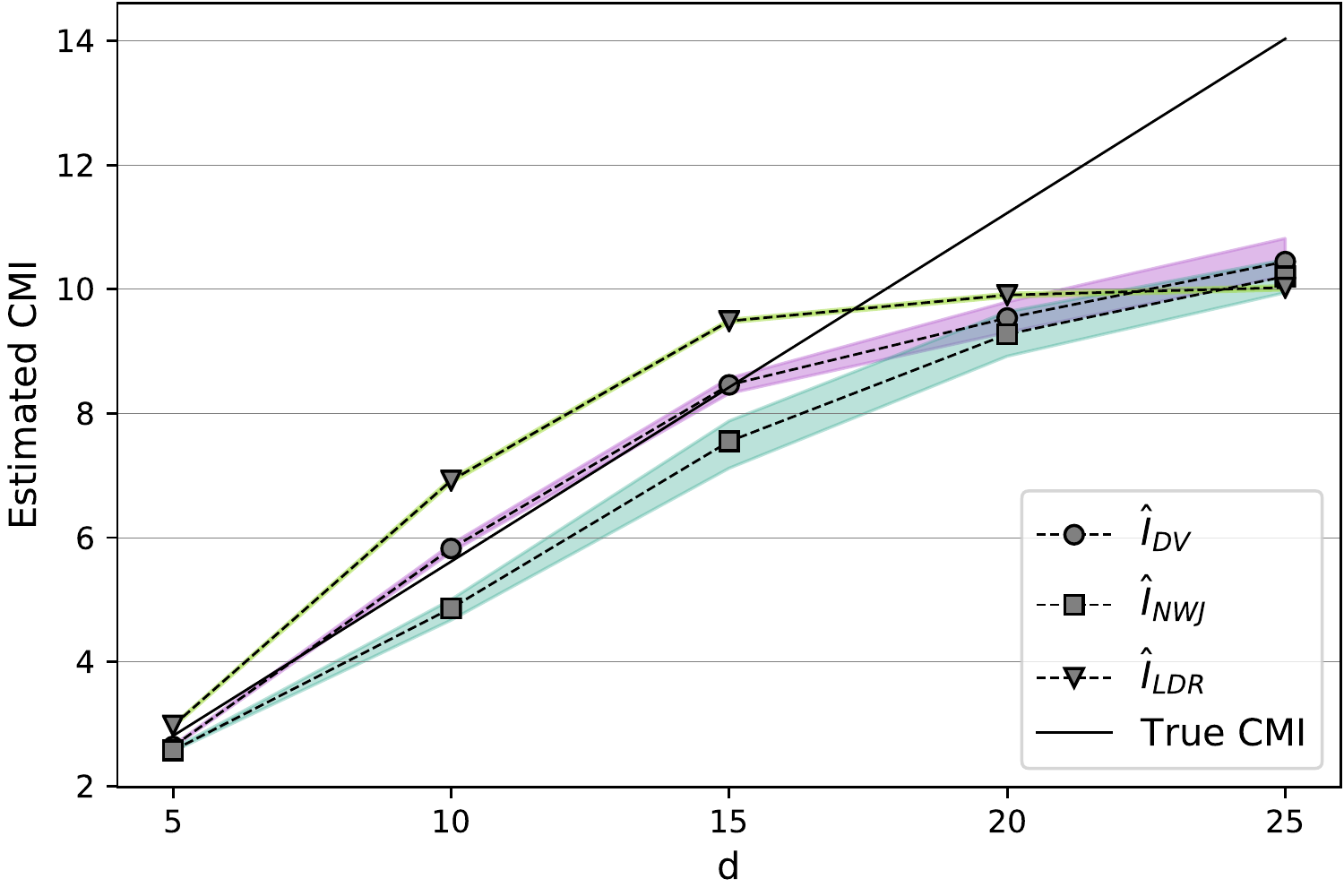}
		\label{fig:dimension_1}
	}\\
	\subfloat[Estimation with $n=1\mathrm{e}{6}$ and $\tau=1\mathrm{e}{-6}$]{
		\includegraphics[width=0.9\linewidth]{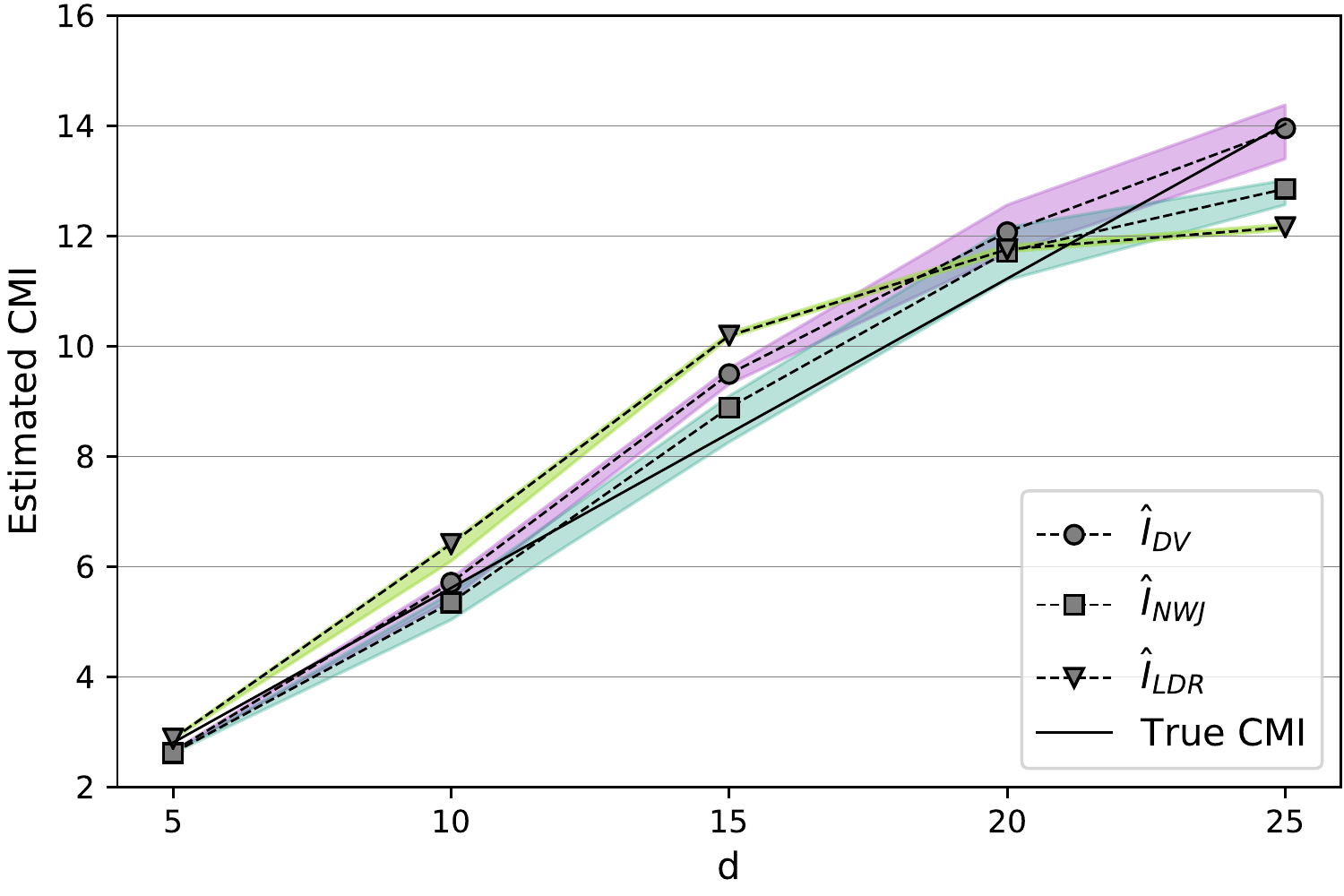}
		\label{fig:dimension_2}
	}	
	\caption{Performance of different estimators using the \textit{isolated $k$-NN} method to estimate $I(X;Y|Z)$ for different data dimensions $d$. The shadows are based on the maximum and minimum values obtained at each iteration.}
	\label{fig:dimension}
\end{figure}

\subsection{Effect of the data dimension}

In order to test the performance of the estimators as the input dimension grows, we consider the generative model~\eqref{eq:model} with $\sigma_x=10$, $\sigma_y=3$, $\sigma_z=5$, and $\Sigma_d=I_d$.
In this experiment, we first estimate the CMI for several values of $d$ using $n=2\mathrm{e}{5}$ number of samples and choosing $\tau=3\mathrm{e}{-5}$.
As depicted in Fig.~\ref{fig:dimension_1}, the estimation seems to saturate for dimension $d=25$. 
This mainly occurs as the output of the neural network is clipped between $[\tau,1-\tau]$, so the density ratio and the estimated CMI are bounded accordingly.

Note that, although a smaller $\tau$ allows the estimator to reach higher values, it also makes the output more unstable (as remarked in~\cite{song2019understanding}). 
To explain this behavior, consider the LDR estimation, $\frac{1}{b}\sum_{\smash{(x,y,z)\in\mathcal{B}^b_\textnormal{joint}}} \log {\hat{\Gamma}(x,y,z)}$.
From~\eqref{eq:Gamma_hat} and assuming $p_1=0.5$, the estimated density ratio $\hat{\Gamma}(x,y,z)$ is bounded by $(1-\tau)/\tau$.
So if $\tau$ is very small, in some rare events, $\hat{\Gamma}(x,y,z)$ can become very large and affect the average.
Nevertheless, by increasing $n$, more typical samples are included and we prevent the odd events from dominating the estimated value.

In Fig.~\ref{fig:dimension_2}, we repeat the experiment with $n=1\mathrm{e}{6}$ and $\tau=1\mathrm{e}{-6}$. It can be seen that the estimations perform more accurately, in particular for higher dimensions. Additionally, it can be noted that the estimated CMI passes over the true value (even based on the NWJ bound) in some cases which is a consequence of choosing a small $\tau$.

\subsection{Non-linear model}

To strengthen our justification on the proposed CMI neural estimator, we consider a non-linear scenario.
First note that for any injective function $f(\cdot)$, $I(f(X);Y|Z)=I(X;Y|Z).$ 
This property allows us to test the performance of the CMI estimators when a non-linear function is applied on the data, while computing the true CMI remains tractable.

\begin{figure}[t]
	\centering
	\includegraphics[width=.9\linewidth]{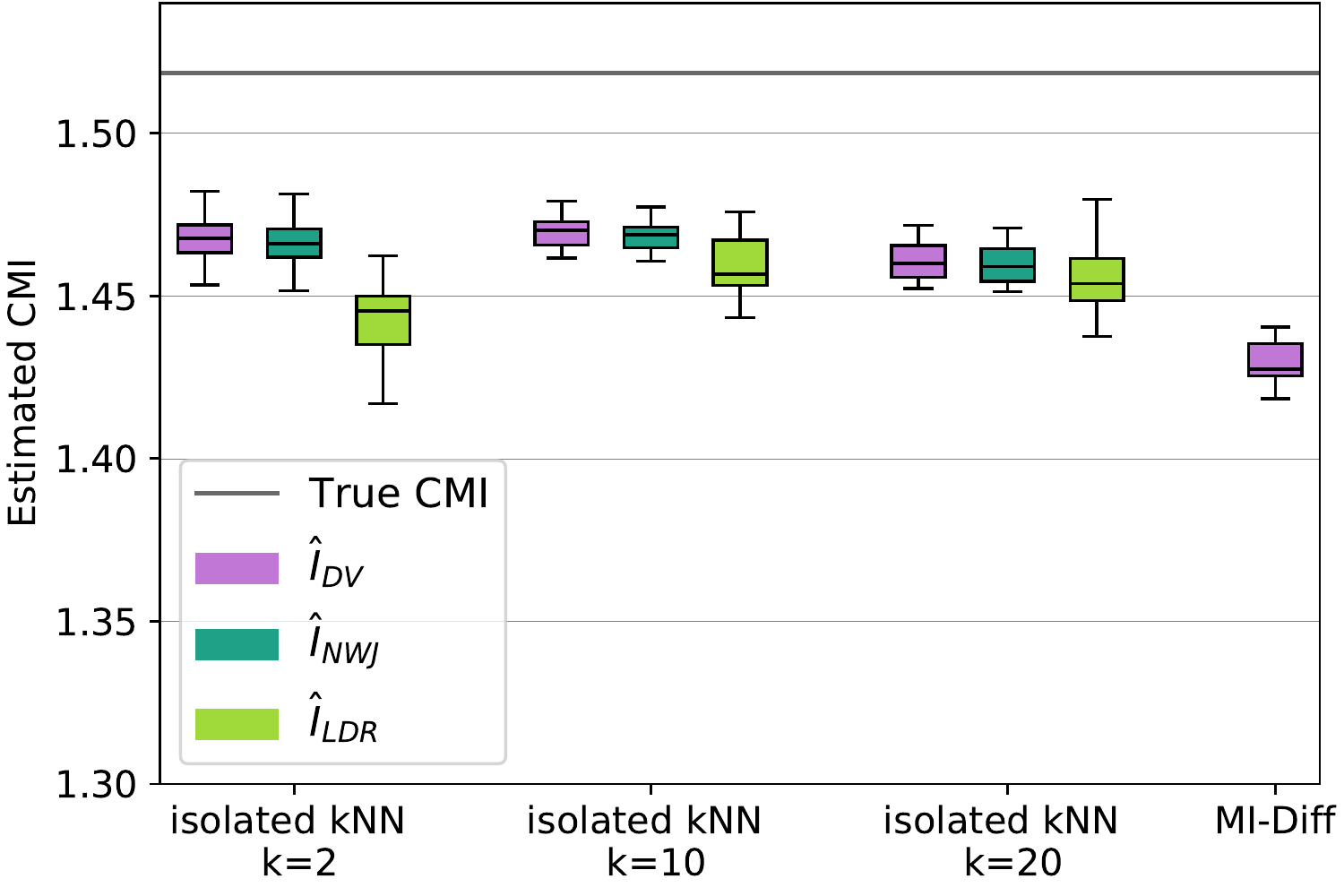}
	\caption{Comparison between the \textit{isolated $k$-NN} and the \textit{MI-Diff} methods to estimate $I(f(X);Y|Z)$, where $f(x)=\tanh(0.05\,x)$, for an input dataset with $n=8\mathrm{e}{4}$ samples and $d=1$.}
	\label{fig:iso-diff_NL}
\end{figure}

We thus estimate $I(f(X);Y|Z)$ with $n=8\mathrm{e}{4}$ samples for $f(x)=\tanh(0.05\,x)$ and the model defined in~\eqref{eq:model}, assuming $\sigma_x$, $\sigma_y$, $\sigma_z$ equal to part~A, while we choose the dimension $d=1$ and $\Sigma_d=1$; the coefficient inside $\tanh(\cdot)$ was chosen to avoid saturation of the output.
The estimation results are plotted in Fig.~\ref{fig:iso-diff_NL}, where it is clear that the non-linear function does not hinder the estimation performance.
To compare our estimators with the \textit{MI-Diff} method, we perform the Mann--Whitney $U$ test.
Since the box plots for the DV and NWJ estimator are superior to the result of the \textit{MI-Diff} method and do not overlap, we only test the LDR estimation results.
We obtained that our LDR estimator performs better than the \textit{MI-Diff} method for all choices of $k$ with $p$-values less than $0.01$.%

\subsection{Additivity and data processing inequality}

In this part, we test two properties of CMI in our estimations.
We consider a slightly more complex model where we assume a component-wise dependency in \eqref{eq:model} by choosing:
\begin{align}
\Sigma_d=\begin{bmatrix}
1 &\rho&0&&&\dots&0\\
\rho&1&\rho&0&&\dots&0\\
0&\rho&1&\rho&0&\dots&0\\
\vdots&&&\vdots&&&\vdots\\
0&&\clap{\dots}&&0&\rho&1
\end{bmatrix}.\nonumber
\end{align} 
Let $Y$, with dimension $d$, be decomposed into two parts $Y_1$ and $Y_2$ with dimensions $d_1$ and $d_2$, respectively. 
Then the conditional mutual information can be split as below:
\begin{align}
I(X;Y|Z)=I(X;Y_1|Z) + I(X;Y_2|Y_1,Z).\label{eq:DPI}
\end{align}
Denote the estimations of $I(X;Y_1|Z)$ and $I(X;Y_2|Y_1,Z)$ by $\hat I_{1,\mathtt{est}}$ and $\hat I_{2,\mathtt{est}}$, respectively.
In this experiment, we test if:
\begin{itemize}
	\item $\hat I_{1,\mathtt{est}} + \hat I_{2,\mathtt{est}}$ can estimate $I(X;Y|Z)$ (Additivity)
	\item $\hat I_{1,\mathtt{est}}$ is smaller than $I(X;Y|Z)$ (data processing inequality (DPI))
\end{itemize}  

\begin{figure}[t]
	\centering
	\includegraphics[width=.95\linewidth]{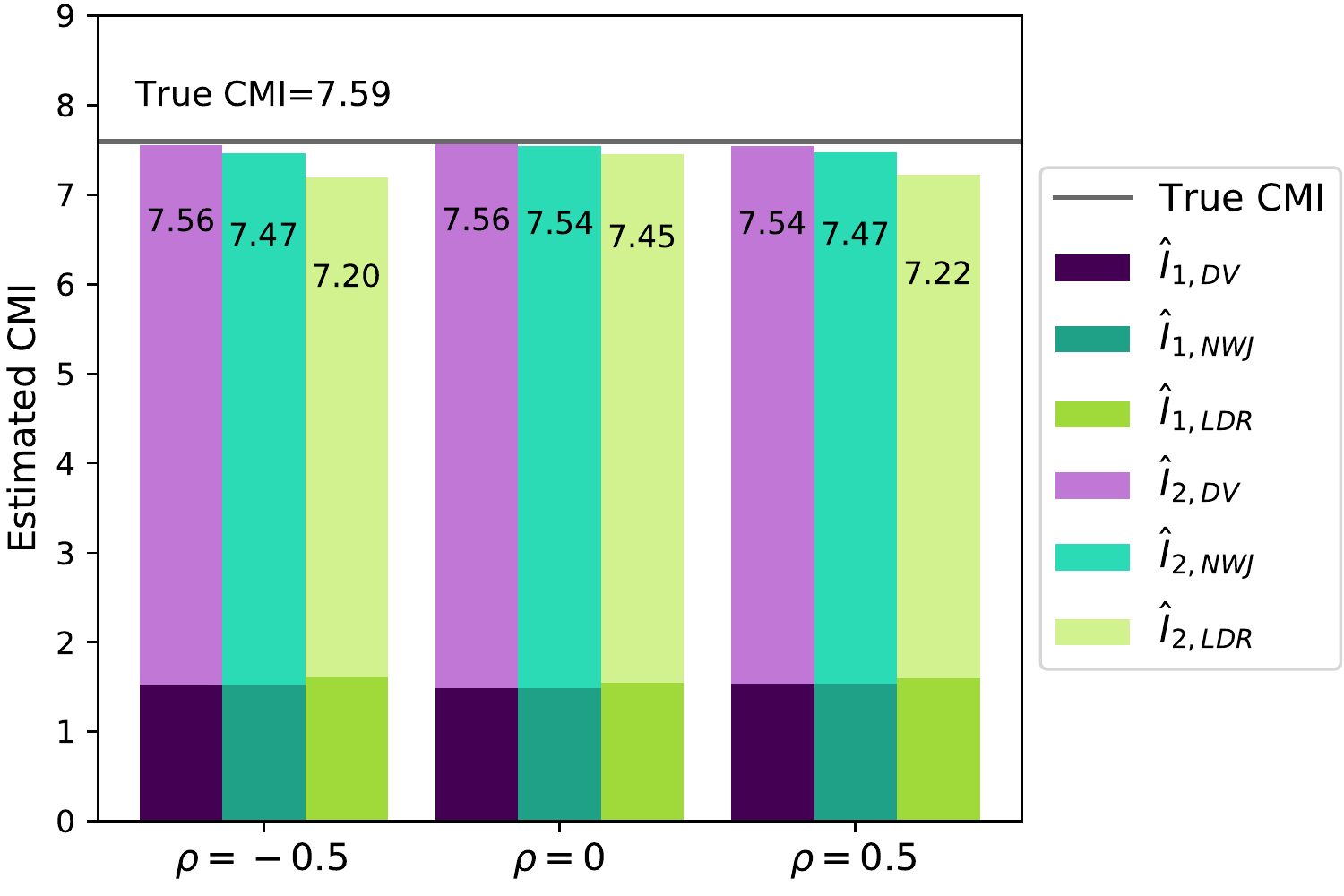}
	\caption{Testing additivity property and DPI for $d=5$, $d_1=1$, and $d_2=4$.}
	\label{fig:DPI}
\end{figure}

Fig.~\ref{fig:DPI} depicts the estimations $\hat I_{1,\mathtt{est}}$ and $\hat I_{2,\mathtt{est}}$ where $\mathtt{est}$ can be replaced with 'DV', 'NWJ', and 'LDR'.
We use $n=2\mathrm{e}{5}$ samples with $d=5$, $d_1=1$, and $d_2=4$, and the results show the averaged estimated values over all trials. 
It can be observed that both the additivity property and DPI hold for different choices of $\rho$.

\begin{figure}[t]
	\centering
	\includegraphics[width=.6\linewidth]{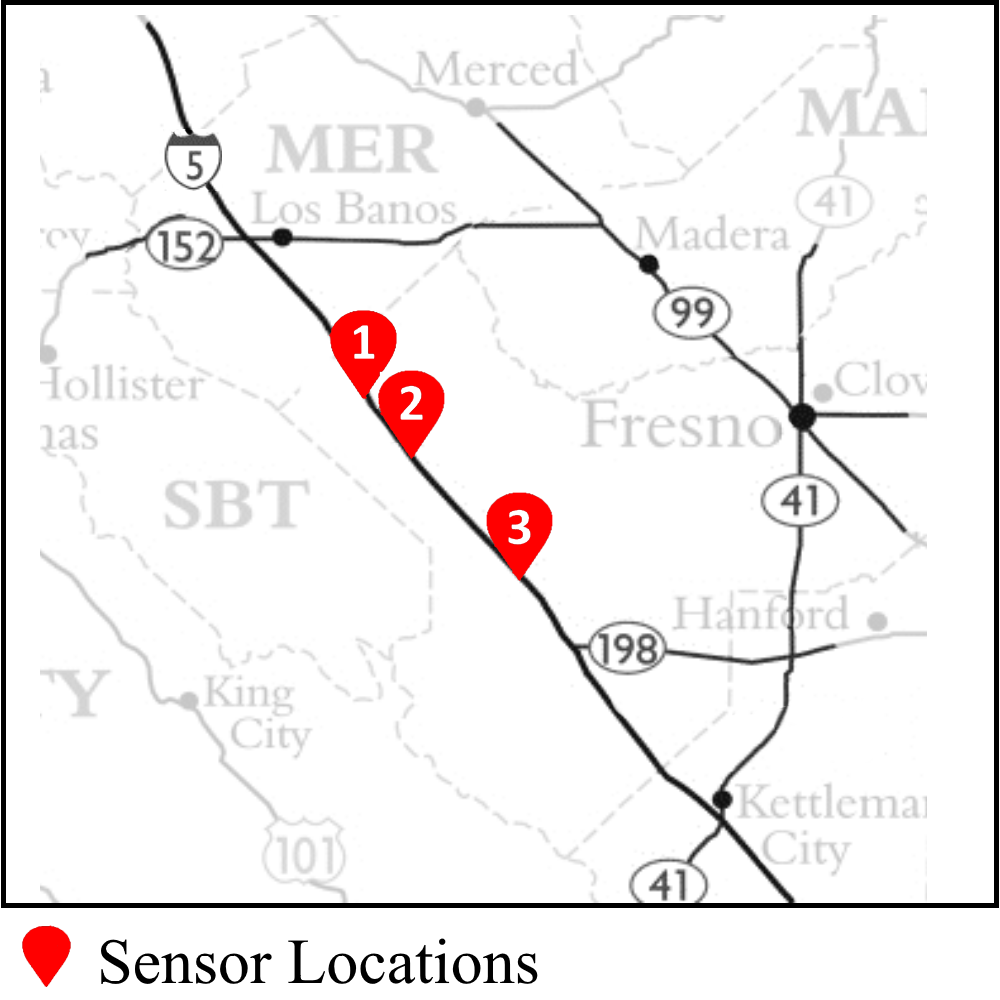}
	\caption{Vehicular traffic sensors mounted on interstate 5 south, California.}
	\label{fig:Fresno}
\end{figure}

\subsection{Real-world scenario}
One application of CMI is in the definition of directed information (DI), which quantifies the extent of causal effects in a network of processes. 
The directed information rate from $X$ to $Y$ causally conditioned on $Z$ is defined as: 
\begin{align*}
I(X\to Y \| Z)\coloneqq\lim_{n\to\infty} \frac{1}{n} \sum_{i=1}^n I(X^i;Y_i|Z^i,Y^{i-1}),
\end{align*}
which if certain Markov properties hold, is simplified as 
\begin{align}\label{eq:DI}
I(X\to Y \| Z)= I(X^l;Y_l|Z^l,Y^{l-1}),
\end{align}
where $l$ is the Markov order in the data model (see~\cite{Mol2017test} for more details).
In other words, directed information captures the dependency of $Y_l$ to the history of $X^l$, when the history of $Y^{l-1}$ and the rest of the network ($Z^l$) is given.
The causal effects are conventionally visualized via a directed graph where the weights of the links are the DI values; this is called a directed information graph (DIG).
Assuming a network with three processes $X$, $Y$, and $Z$,  the weight of the link $X\to Y$ in the DIG represents $I(X\to Y \| Z)$.   

In~\cite{zhang2019itene}, the authors exploited the neural estimator in~\cite{mukherjee2019ccmi_conf} to estimate the transfer entropy, which is a similar notion to DI.
This motivates applying our neural estimators for DI, although the data is not i.i.d. but rather Markov.
In this experiment, we apply our estimators on a real-world dataset which is collected from vehicular traffic sensors mounted on California highways\footnote{Accessed from California Department of Transportation (\url{http://pems.dot.ca.gov/}).}.
Estimating DIG for vehicular traffic has been studied in~\cite{molavipourTraffic} by quantizing the measurements' values and exploiting classical methods for discrete-value processes.

In this experiment, traffic flow data represents the number of cars passing the sensor for a time period of 5 minutes. 
We choose three consecutive sensors on the interstate 5 south, in Fresno (see Fig.~\ref{fig:Fresno}), and the data is aggregated from January 2017 to June 2019, every day, for 24 hours.
From the approximate distance between sensors and the average speed of the cars, we choose $l=5$ in~\eqref{eq:DI} to estimate the DI rate and construct the DIG graph. 
Since the sensors are in a row, the vehicles which are observed in one sensor will appear in the next sensor with a delay.
However, due to the inbound and outbound traffic, it is possible to miss part of the flow.

The estimated DI between sensors is stated below in terms of weights of the adjacency matrix of the DIG:
\begin{align*}
G=\begin{bmatrix}
0 & 0.22 & 0.06\\
0.1 & 0 & 0.1\\
0.01 & 0.06 & 0
\end{bmatrix}.
\end{align*}
The results indicate that the most significant causal effects are between sensor~1 and sensor~2, and from sensor~2 to sensor~3.
This is in line with the physical placement of the sensors.
Note that in~\eqref{eq:DI} the dependency of $X_l$ on $Y_l$ is included in the DI.
So if a car passes two sensors in the same time interval, it will be considered in both traffic flow signals.
The backward link from sensor~2 to sensor~1 could be the result of this event since they are closer in distance (see Fig.~\ref{fig:Fresno}).

\section{Discussion and Final Remarks}
\label{sec:conclude}

In this paper, we studied the use of a neural network classifier to estimate the conditional mutual information among some random variables, based on a dataset composed of i.i.d. samples of said random variables. 
Inspired by the $k$-NN method, we introduced a new technique for creating sample batches which re-samples the existing dataset; this re-sampling models a particular conditional independence in the distribution of the new samples.
The classifier is then trained to distinguish between the original distribution of samples and the new one.
This technique enabled us to estimate the CMI directly rather than estimating it as the difference of two MI terms. 
Our simulations showed that estimating with the proposed \textit{isolated $k$-NN} method improved the accuracy of estimation for high and low values of CMI in several scenarios.
However, extending the results to more complicated models or other real-world data requires further investigation in tuning the neural network and choosing its activation functions.
This can be considered as a future direction of this work.

While the estimation based on the NWJ bound is always less than the DV estimation, we cannot make a general claim about the LDR estimation.
For instance, in Fig.~\ref{fig:iso2} the LDR estimation is higher than the DV estimation, while this does not hold in the non-linear experiment (Fig.~\ref{fig:iso-diff_NL}) or the results in Fig.~\ref{fig:DPI}.
In addition, we emphasize that a higher estimation does not necessarily reduce the bias as the estimations may overshoot the true value; for example in Fig.~\ref{fig:iso-diff_zero} or Fig.~\ref{fig:dimension_2}.

Neural networks have been proposed in communication systems as part of the encoder/decoder blocks~\cite{o2017introduction, dorner2017deep}.
However, learning the end-to-end communication system requires knowing the channel model, which might not be available in practice.
While there exist approaches based on generative adversarial networks (GAN) \cite{o2018physical}, in~\cite{fritschek2019deep}, the authors optimize channel encoders by estimating the MI and advocate the use of neural estimators.
This approach can be followed with our proposed estimators for channels with capacities characterized by CMI.
However, as noted in~\cite{molavipour2020conditional}, one should be careful to use an appropriate estimator for CMI; if the estimated value is used to determine the transmission rate and it is above the true value of CMI, the system will experience a catastrophic failure.

In some applications, a key step is to perform a threshold test on CMI rather than estimating its exact value.
For instance in~\cite{Mol2017test}, the causal links of a network of random processes can be detected by checking if the CMI is above a certain threshold.
In order to achieve a high accuracy in such tests (i.e., small type-I/II errors), it is not necessarily required to estimate the CMI accurately.
Therefore, the performance of the tests can be investigated as a future direction of this work.

\appendices
\section{Proof of Theorem~\ref{th:convergence}}
\label{App:proof_th_convergence}

\subsection{Preliminaries}

First let us review the lemmas that we require in this proof.
Since each of the terms in $\hat{g}(x^n,y^n,z^n)$ is bounded, McDiarmid's inequality \cite{mcdiarmid1989method} is exploited to obtain concentration bounds.

\begin{lemma}[McDiarmid's inequality]
	\label{Lemma:McDiarmid}
	Let $V_1,\dots,V_n$ be independent random variables $V_i\in\mathcal{V}$ and assume $\phi:\mathcal{V}^n\to\mathbb{R}$ such that for all $i\in\{1,\dots,n\}$:
	$$\sup_{\substack{v_1,\dots,v_n \\ v'_i}} \abs{\phi(v_1,\dots,v_n)-\phi(v_1,\dots,v'_i,\dots,v_n)}\leq c_i.$$
	Then the following bound holds:
	$$\pr{\abs{\phi(V^n)-\ex[\phi(V^n)]}\geq \epsilon}\leq 2\exp\left(\frac{-2\epsilon^2}{\sum_{i=1}^{n}c_i^2}\right).$$
\end{lemma}

The inner sum in $\hat g(\cdot)$, defined in \eqref{eq:g_hat_def}, resembles a $k$-NN regression which we leverage in our proof.

\begin{lemma}[{\cite[Theorem~1]{devroye1994strong}}]
	\label{Lemma:L1_kNN}
	Let $(U_1,Z_1),\dots,(U_n,Z_n)$ be generated i.i.d. according to $p(u,z)$ and let $\zeta\in\mathcal{X}$.
	Using the same notation as in Definition~\ref{def:isoKnn}, let $\mathcal{A}^m(\zeta)$ be the set of indices of the elements of $z^n$ which are the $k$-NN of $\zeta$, and define
	\begin{equation*}
	\psi^m_n(\zeta)\coloneqq\frac{1}{k}\sum\nolimits_{j\in\mathcal{A}^m(\zeta)} U_j,
	\end{equation*}
	and $\bar\psi(\zeta)\coloneqq\ex_{p(u|z)}[U|Z=\zeta]$.
	Further assume that $\abs{U}\leq M$ and $\lim_{n\to\infty}k(n)=\infty$ and $\lim_{n\to\infty}k(n)/n=0$.
	Then, if the neighbors are chosen in $Z_1^n$ (i.e., by taking indices in $\mathcal{A}^0(\cdot)$), for any $\epsilon>0$ there exists an integer $n_0$ such that for $n>n_0$:
	\begin{align}
	\label{eq:psi_bnd}
	\pr{\int p(z)\abs{\psi^0_n(z)- \bar\psi(z)}dz >\epsilon} \leq \exp\! \left(\frac{-n\epsilon^2}{8 M^2\gamma_d^2}\right)\!,
	\end{align}
	where $\gamma_d$ is the minimal number of cones centered at the origin, of angle $\pi/6$, that cover $\mathbb{R}^d$.
\end{lemma}

\begin{remark}\label{remark:L1_kNN}
	In Lemma~\ref{Lemma:L1_kNN}, according to the definition of $\mathcal{A}^m(\zeta)$, we assume that $m=0$.
	Nevertheless, if $0<m<n$ and $\lim_{n\to\infty}k(n)/{(n-m)}=0$, since the pairs are i.i.d., similar to \eqref{eq:psi_bnd}, we have that
	\begin{equation*} 
	\pr{\int p(z) \Big| \psi^m_n(z)- \bar\psi(z) \Big| {dz} >\epsilon} \leq \exp\! \left(\frac{-(n-m)\epsilon^2}{8 M^2\gamma_d^2}\right)\!.
	\end{equation*}
\end{remark}

\subsection{Proof of Theorem~\ref{th:convergence}}
To begin the proof, we make the following definitions:
\begin{align*}
g^k_n(y,z) &\coloneqq \frac{1}{k}\sum\nolimits_{j\in\mathcal{A}^m(z)} g(x_j,y,z),\\
\bar{g}(y,z) &\coloneqq \ex_{p(x|z)}\big[ g(X,y,z) \big].
\end{align*}
Note that $g^k_n(y,z)$ is in fact a function of $x^n,z^n$ and $y,z$; We use the simplified notation as the dependence on the data can be understood from the context.
Moreover, since the pairs $(y_i,z_i)$ are i.i.d. from the dataset, we may assume $\mathcal{I}_{m}=\{1,\dots,m\}$ without loss of generality.
We thus rewrite the estimator~\eqref{eq:g_hat_def} as: 
\begin{align}
\hat g(x^n,y^n,z^n)=\frac{1}{m}\sum_{i=1}^{m} g^k_n(y_i,z_i).
\end{align}
Now, using the triangle inequality, we have that
\begin{align}
\MoveEqLeft
\abs{\hat g(x^n,y^n,z^n)-\ex_{\dprod}[g(X,Y,Z)]}\nonumber\\
&= \Bigg| \frac{1}{m}\sum_{i=1}^m g^k_n(y_i,z_i) - \int p(y,z)\bar{g}(y,z)\, dy\, dz \Bigg| \nonumber\displaybreak[2]\\
&\leq \Bigg| \frac{1}{m}\sum_{i=1}^m g^k_n(y_i,z_i) - \int p(y,z)g^k_n(y,z)\, dy\, dz \Bigg| \nonumber\\
&\quad +\abs{\int p(y,z)\big(g^k_n(y,z)-\bar{g}(y,z)\big) dy\, dz}\nonumber\\
&\leq \Bigg| \frac{1}{m}\sum_{i=1}^m g^k_n(y_i,z_i) - \int p(y,z)g^k_n(y,z)\, dy\, dz \Bigg| \nonumber\\ 
&\quad +\int p(z)\abs{\int p(y|z)\big(g^k_n(y,z) - \bar{g}(y,z)\big)dy} dz\nonumber\\
&\leq \Bigg| \frac{1}{m}\sum_{i=1}^m g^k_n(y_i,z_i) - \ex \big[ g^k_n(Y,Z) \big] \Bigg| \nonumber\\
&\quad +\abs{\int p(y,z)g^k_n(y,z)\, dy\, dz-\ex \big[ g^k_n(Y,Z) \big]}\nonumber\\
&\quad +\int p(z)\abs{\int p(y|z)\big(g^k_n(y,z) - \bar{g}(y,z)\big)dy} dz. \label{eq:tri_bnd}
\end{align}
To elaborate on the first two terms on the RHS of~\eqref{eq:tri_bnd}, we note that $\frac{1}{m}\sum_{i=1}^m g^k_n(y_i,z_i)$ is a function of the random variables $X_{m+1}^n,Y^m,Z^n$, and thus random itself, while the randomness of $\int p(y,z)g^k_n(y,z)\, dy\, dz$ in the second term stems from $X_{m+1}^n,Z_{m+1}^n$. 
On the other hand, $\ex[g^k_n(Y,Z)]$ is a deterministic term and the expectation is with respect to the density function $p(y,z)p(x_{m+1}^n,z_{m+1}^n)$.

In the following, we show the convergence of the first two terms in~\eqref{eq:tri_bnd} according to Lemma~\ref{Lemma:McDiarmid}. Next we show that the last term converges to zero according to Lemma~\ref{Lemma:L1_kNN} and Remark~\ref{remark:L1_kNN}. Note that Assumption~\ref{Assum:knn} guarantees the required assumption on $k$ in Lemma~\ref{Lemma:L1_kNN} and Remark~\ref{remark:L1_kNN}.

\subsubsection{First term \texorpdfstring{in~\eqref{eq:tri_bnd}}{}}
Define $w_i\coloneqq(x_i,y_i,z_i)$ and let
\begin{equation*}
\phi\left( w^n\right) = \frac{1}{m}\sum_{i=1}^m g_n^k(y_i,z_i), 
\end{equation*}
which is a function of the random triples $\left\{(X_i,Y_i,Z_i)\right\}_{i=1}^n$. 
For any $i\in\{1,\dots,n\}$ we have that
\begin{multline}\label{eq:phi_bnd_diff}
\smash{ \sup_{\substack{w^n, w'_i}} } \abs{\phi(w_1,\dots,w_n) - \phi(w_1,\dots,w'_i,\dots,w_n)}\\ \leq \sm{\frac{c}{\min\{m,k\}}}, 
\end{multline} 
where \sm{$c \coloneqq g^{max}-g^{min}$}.
To see this, first consider a triple $(x_i,y_i,z_i)$ is altered to $(x'_i,y'_i,z'_i)$ for $i\leq m$.
Then the largest difference that can happen is \sm{$c/m$}.
In case $i>m$, the extreme case is that $z_i$ is the neighbor of all $z_1,\dots,z_m$, so in total the difference becomes \sm{$c/k$}.
By Assumption~\ref{Assum:knn}, $m>k$ and thus \sm{the RHS of \eqref{eq:phi_bnd_diff} becomes $c/k$}.

Since~\eqref{eq:phi_bnd_diff} holds, Lemma~\ref{Lemma:McDiarmid} implies the following bound:
\begin{equation*}
\pr{\abs{\phi(W^n) - \ex_{p(w^n)}\big[\phi(W^n)\big]} >\epsilon}\leq \sm{2\exp\left(\frac{-2\epsilon^2 k^2}{nc^2}\right)}.
\end{equation*}
The expectation inside the left hand side (LHS) of this equation may be rewritten as follows:
\begin{align}
&\ex_{p(w^n)} \big[ \phi\left(W^n\right) \big] = \frac{1}{m}\ex_{p(w^n)} \Bigg[\sum_{i=1}^{m} g_n^k(Y_i,Z_i)\Bigg] \nonumber\\
&\quad= \frac{1}{m}\sum_{i=1}^{m} \ex_{p(y_i,z_i)p(x_{m+1}^n,z_{m+1}^n)} \Big[ g^k_n(Y_i,Z_i) \Big] \nonumber\\
&\quad= \ex_{p(y,z)p(x_{m+1}^n,z_{m+1}^n)} \Big[ g^k_n(Y,Z) \Big],
\end{align}
where the last equality holds since the pairs $(Y_i,Z_i)$ are generated i.i.d.
As a result,
\begin{multline}
\label{eq:McD_3}
\mathds{P}\Bigg(\bigg\lvert\frac{1}{m}\sum_{i=1}^{m} g_n^k(Y_i,Z_i) - \ex\big[ g^k_n(Y,Z) \big]\bigg\rvert >\epsilon\Bigg) \\[-1mm]
\leq \sm{2\exp\Bigg(\frac{-2\epsilon^2 k^2}{nc^2}\Bigg)}.
\end{multline}

\subsubsection{Second term \texorpdfstring{in~\eqref{eq:tri_bnd}}{}}

Similarly, let 
\begin{equation}
\phi'(w_{m+1}^n)=\int p(y,z)g^k_n(y,z)\, dy\, dz;
\end{equation}
then, for any $i\in\{m+1,\dots,n\}$, we have that
\begin{align*}
&\sup_{\substack{w_{m+1}^n, w'_i}} \abs{\phi'(w_{m+1}^n) - \phi'(w_{m+1},\dots,w'_{i},\dots,w_n)} \leq \sm{\frac{c}{k}}.
\end{align*}
Hence, Lemma~\ref{Lemma:McDiarmid} yields the following bound:
\begin{multline}
\label{eq:McD2}
\pr{\abs{\phi'\big(W_{m+1}^n\big) - \ex\big[ \phi' \big( W_{m+1}^n \big)\big]} >\epsilon}\\
\leq \sm{2\exp\left(\frac{-2\epsilon^2 k^2}{(n-m)c^2}\right)}.
\end{multline}
The deviation of the second term in~\eqref{eq:tri_bnd} can thus be bounded as below:
\begin{align}
\MoveEqLeft[2]
\pr{\abs{\int p(y,z)g^k_n(y,z)\, dy\, dz - \ex\big[ g^k_n(Y,Z) \big]}>\epsilon}\nonumber\displaybreak\\
&=\begin{multlined}[t]
\mathds{P}\Bigg(\bigg\lvert\int p(y,z)g^k_n(y,z)\, dy\, dz\\
-\ex_{p(y,z)p(x_{m+1}^n,z_{m+1}^n)} \big[ g^k_n(Y,Z) \big]\bigg\rvert>\epsilon\Bigg)
\end{multlined}\nonumber\\
&=\pr{\abs{\phi'(W_{m+1}^n) - \ex\big[\phi'(W_{m+1}^n)\big]} >\epsilon}\nonumber\\
&\leq \sm{2\exp\left(\frac{-2\epsilon^2 k^2}{(n-m)c^2}\right)},\label{eq:McD2_2}
\end{align}
where the last step is due to~\eqref{eq:McD2}.

\subsubsection{Third term \texorpdfstring{in~\eqref{eq:tri_bnd}}{}}

Note that, for any $z$ and any $j\in\mathcal{A}^m(z)$, and given the assumption of the theorem,
\begin{equation*}
\abs{\int p(y|z)g(x_j,y,z) dy}\leq M. 
\end{equation*}
So if Assumption~\ref{Assum:knn} holds, we know from Lemma~\ref{Lemma:L1_kNN} that, for any $\epsilon>0$, there exists an integer $n_0$ such that for $n>n_0$
\begin{multline}\label{eq:KNN_third_term}
\pr{\int p(z)\abs{\int p(y|z)\big(g^k_n(y,z) - \bar{g}(y,z)\big)dy} dz >\epsilon} \\
\leq \exp \left(\frac{-(n-m)\epsilon^2}{8 M^2\gamma_d^2}\right).
\end{multline}
Therefore, combining~\eqref{eq:tri_bnd}, \eqref{eq:McD_3}, \eqref{eq:McD2_2}, and~\eqref{eq:KNN_third_term}, we have that%
\begin{multline*}
\prB{\abs{\hat g(x^n,y^n,z^n)-\ex_{\dprod}\big[g(X,Y,Z)\big]}\geq 3\epsilon}\\
\leq\mdel{1}(\epsilon,c,M).
\end{multline*} 
This concludes the proof of Theorem~\ref{th:convergence}.
\hfill$\square$

\input{Appendix_B}

\input{Appendix_C}

\input{Appendix_F}

\input{Appendix_G}

\ifCLASSOPTIONcaptionsoff
  \newpage
\fi

\bibliographystyle{IEEEtran}
\bibliography{IEEEabrv,ref}

\end{document}

%% file: pack.tex
\usepackage[utf8]{inputenc}
\usepackage{amsmath, amssymb, amsthm, commath, mathtools,amsfonts}
\usepackage{tikz}
\usetikzlibrary{calc}
\usetikzlibrary{shapes,arrows,fit,patterns,positioning,decorations.markings,automata}
\usepackage{graphicx}
\usepackage[noadjust]{cite}
\usepackage{algorithmic}
\usepackage{textcomp}
\usepackage{xcolor}
\usepackage{dsfont}
\usepackage[caption=false]{subfig}
\usepackage[ruled,vlined,linesnumbered]{algorithm2e}

\newcommand{\ex}[0]{\mathds{E}}
\newcommand{\pr}[1]{\mathds{P}\!\left ( #1 \right)}
\newcommand{\prB}[1]{\mathds{P}\bigg ( #1 \bigg)}
\newcommand{\klD}[2]{D\!\left(#1\,||\,#2\right)}

\newcommand{\myabs}[1]{\bigg\lvert#1\bigg\rvert}

\renewcommand\thesubsubsection{\Alph{subsubsection}}

\newtheorem{definition}{\bf Definition}
\newtheorem{assumption}{\bf Assumption}
\newtheorem{theorem}{\bf Theorem}
\newtheorem{lemma}{\bf Lemma}
\newtheorem{remark}{\bf Remark}

\newtheorem{corollary}{\bf Corollary}
\newtheorem{proposition}{\bf Proposition}

\newcommand{\DV}[0]{\textit{DV}}
\newcommand{\NWJ}[0]{\textit{NWJ}}
\newcommand{\LDR}[0]{\textit{LDR}}
\newcommand{\emp}[0]{\textnormal{emp}}

\newcommand{\vnorm}[1]{\| #1 \|}

\newcommand{\djoint}[0]{p(x,y,z)}
\newcommand{\dprod}[0]{p(x|z)p(y,z)}

\newcommand{\mdel}[1]{\delta_{#1}^n}

%% file: net_fig.tex
\begin{figure}
	\centering

\resizebox{0.6\linewidth}{!}{%
\begin{tikzpicture}

\tikzset{
	input/.style={
	draw=gray!10,
	rectangle,
	rounded corners=5pt,
	align=center,
	fill=gray!10
	},
	data/.style={
	draw=gray!10,
	rounded rectangle,
	fill=gray!10
	},
	dot/.style={
	circle,
	draw=gray, 
	fill=gray, 
	inner sep=0pt,
	minimum size=3pt
	},
	Neu/.style={
	circle,
	draw=gray!30, 
	fill=gray!30, 
	inner sep=0pt,
	minimum size=.5cm
	},
	mLink/.style={
	gray,
	line width=.1pt	
	},
	sig path/.style={
	path picture={
		\node at (path picture bounding box.center) {
			\includegraphics[height=0.2cm]{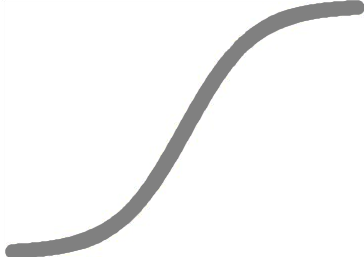}};}}
}

\node (xi1) [input,text width=.2cm, font=\scriptsize] at (0.5,-0.8) { X \\ \vspace{.3cm} Y \\ \vspace{.3cm} Z};

\node (N1) [Neu] at (2,0.25) {};
\node (N2) [Neu] at (2,-0.45) {};
\node (N3) [Neu] at (2,-1.15) {};
\node (N4) [Neu] at (2,-1.85) {};

\node (M1) [Neu] at (3,0.25) {};
\node (M2) [Neu] at (3,-0.45) {};
\node (M3) [Neu] at (3,-1.15) {};
\node (M4) [Neu] at (3,-1.85) {};

\node (S) [Neu] at (4,-0.8) {};

\node (O) [	circle,draw=gray!30,inner sep=0pt,minimum size=.5cm,sig path] at (5,-0.8) {};

\node (t) at (5.7,-0.8) {$\omega_{\theta}$};

\draw[mLink] (1,-0.2) to (N1);
\draw[mLink] (1,-0.2) to (N2);
\draw[mLink] (1,-0.2) to (N3);
\draw[mLink] (1,-0.2) to (N4);

\draw[mLink] (1,-1.4) to (N1);
\draw[mLink] (1,-1.4) to (N2);
\draw[mLink] (1,-1.4) to (N3);
\draw[mLink] (1,-1.4) to (N4);

\draw[mLink] (M1) to (N1);
\draw[mLink] (M1) to (N2);
\draw[mLink] (M1) to (N3);
\draw[mLink] (M1) to (N4);

\draw[mLink] (M2) to (N1);
\draw[mLink] (M2) to (N2);
\draw[mLink] (M2) to (N3);
\draw[mLink] (M2) to (N4);

\draw[mLink] (M3) to (N1);
\draw[mLink] (M3) to (N2);
\draw[mLink] (M3) to (N3);
\draw[mLink] (M3) to (N4);

\draw[mLink] (M4) to (N1);
\draw[mLink] (M4) to (N2);
\draw[mLink] (M4) to (N3);
\draw[mLink] (M4) to (N4);

\draw[mLink] (M1) to (S);
\draw[mLink] (M2) to (S);
\draw[mLink] (M3) to (S);
\draw[mLink] (M4) to (S);

\draw[mLink] (S) to (O);

\end{tikzpicture}
}
\caption{The proposed neural network classifier that is parameterized with $\theta$ and consists of two hidden layers and concatenated with a sigmoid function.}
\label{fig:net}
\end{figure}

%% file: Appendix_B.tex
\section{Proof of Proposition~\ref{Prop:convergence_L}}
\label{App:proof_prop_L}

The proof combines a concentration bound for $L_b^1(\omega_\theta)$, which follows from Hoeffding's inequality, and a concentration bound for $L_{b'}^2(\omega_\theta)$, which follows from Theorem~\ref{th:convergence}.

By construction, $\mathcal{B}_\textnormal{joint}^b$ consists of i.i.d. samples distributed according to $p(x,y,z)$. 
Moreover, the summands in~\eqref{eq:Lb12} are bounded, i.e.,
\begin{equation*}
\log\tau\leq\log\omega_{\theta}(x,y,z)\leq\log(1-\tau),
\end{equation*}
given that the output of the classifier is clipped.
Therefore, Hoeffding's inequality may be directly applied to find a concentration bound on $L_b^1(\omega_\theta)$, as seen in the lemma below.

\begin{lemma}\label{Lemma:Lb1}
	For all $\theta$ and given $\epsilon>0$, the following inequality holds:
	\begin{multline}
	\pr{\abs{L_b^1(\omega_\theta)+\ex_{p(x,y,z)} \big[\log \omega_{\theta}(X,Y,Z)\big]}\geq \epsilon}\nonumber\\
	\leq 2\exp\left(-\frac{2b\epsilon^2}{(\log\frac{1-\tau}{\tau})^2}\right).
	\end{multline}
\end{lemma}

Next, we show the convergence of $L_{b'}^2(\omega_\theta)$. 
According to Theorem~\ref{th:convergence}, \eqref{eq:Lb12}, and the definition of $\mathcal{B}^{b'}_\textnormal{prod}$, if we consider the function $g(x,y,z)=-\log \big(1-\omega_\theta(x,y,z) \big)$, we have that $\hat g(x^n,y^n,z^n)=L_{b'}^2(\omega_\theta)$.
In this case, $g^{max}=-\log(\tau)$, $g^{min}=-\log(1-\tau)$, and $M=-\log(\tau)$, which implies that \sm{$c=\log\frac{1-\tau}{\tau}$}.
Then, the following Corollary is deduced.

\begin{corollary}\label{Cor:Lb2}
	Let Assumption~\ref{Assum:knn} hold, then for any $\theta$ there exists $n_0$ such that for $n>n_0$:
	\begin{multline*}
	\!\!\pr{\abs{L_{b'}^2(\omega_\theta)+\ex_{p(x|z)p(y,z)}\big[ \log \big(1- \omega_\theta(X,Y,Z) \big) \big]}> 3\epsilon}\\
	\leq \mdel{2}(\epsilon),
	\end{multline*}
	where $\mdel{2}(\epsilon)$ is defined in Table~\ref{tab:param}. 
\end{corollary}

Now by the triangle inequality, we have that
\begin{align}
\MoveEqLeft[0]
\abs{L_{\emp}(\omega_\theta)-L(\omega_\theta)} \nonumber\\
&\leq (1-p_1) \abs{L_b^2(\omega_\theta)+\ex_{p(x|z)p(y,z)} \big[\log (1-\omega_{\theta}(X,Y,Z)) \big]} \nonumber\\
&\quad +p_1\abs{L_b^1(\omega_\theta)+\ex_{p(x,y,z)} \big[\log \omega_\theta(X,Y,Z)\big]},
\end{align}
and the proof of Proposition~\ref{Prop:convergence_L} is complete by combining Lemma~\ref{Lemma:Lb1}, Corollary~\ref{Cor:Lb2}, and choosing $\epsilon=\frac{\mu}{3-2p_1}$.
\hfill$\square$

%% file: Appendix_C.tex
\section{Proof of Theorem~\ref{th:consistency}}
\label{appendix:Proof_Th2}

The consistency of our proposed estimators is tied to the approximation power of the neural network, which is addressed in the following lemma.

\begin{lemma}
	\label{Lemma:functiona_apprx}
	For a given $\epsilon>0$, $\exists\, \tilde\theta\in\Theta$ such that $\Theta\subset \mathbb{R}^h$ is compact and
	$$\abs{L(\omega_{\tilde\theta})-L^*}\leq \frac{\epsilon}{2}.$$
	\begin{proof}
		The proof can be shown similar to \cite[Lemma~4]{mukherjee2019ccmi_conf}.
	\end{proof}
\end{lemma}

\begin{remark}
	The universal functional approximation introduced in \cite{hornik1989multilayer} allows choosing parameters in a compact set of $\mathbb{R}^h$, while $\epsilon$ determines the number of neurons, and accordingly $h$, such that the desired approximation is achieved.
	Consider the network is approximating the function $\omega^*$; then, given $\epsilon>0$, there exist a set $\Theta$ and a parameter $\tilde{\theta}\in\Theta$ such that $\omega_{\tilde{\theta}}$ is at an $\epsilon$ distance of $\omega^*$.
\end{remark}

Adopting an optimizer such as \textit{Adam}, we can minimize $L_{\emp}(\omega_\theta)$ to find $\hat\theta$, and it is desired that $L(\omega_{\hat{\theta}})$ is close to $L^*$, which suggests we can use the neural network to approximate $\omega^*$.
As shown in Proposition~\ref{Prop:convergence_L}, given a particular $\theta\in\Theta$, one can choose $n$ such that $L_{\emp}(\omega_\theta)$ falls in a the neighborhood of $L(\omega_\theta)$.
Nevertheless, we need a more restrictive condition if we want to guarantee such convergence for all $\theta$ simultaneously.
This is addressed in the lemma below. 

\begin{lemma}\label{Lemma:convergence_Sup_L}
	Let Assumptions~\ref{Assum:knn} and \ref{Assum:NN_Lip} hold, then for any $\mu>0$, there exists $n_1$ such that, for $n>n_1$, we have that
	\begin{equation}
	\pr{\sup_{\theta\in \Theta} \abs{L_{\emp}(\omega_\theta)-L(\omega_\theta)}>2\mu} \leq \mdel{4}(\mu),
	\label{eq:lemma_covering}
	\end{equation}
	where $\mdel{4}$ is defined in Table~\ref{tab:param}.
	\begin{proof}
		The proof follows similar steps as the one for~\cite[Lemma~5]{mukherjee2019ccmi_conf} and we provide here only some details.
		Since $\Theta\subset \mathbb{R}^h$ and $\vnorm{\theta}_2\leq K$, $\forall\theta\in\Theta$, $\Theta$ can be covered with $N(\Theta,r)$ number of balls of radius $r$---the covering number with respect to $\ell_2$.
		The covering number is finite and bounded \cite{shalev2014understanding}:
			\begin{align}\label{eq:covering_bound}
			N(\Theta,r)\leq \bigg( \frac{2K\sqrt{h}}{r} \bigg)^h.
			\end{align}
		Let $\{ \theta_1, \dots, \theta_{N(\Theta,r)} \}$ denote the centers of the covering balls and $\{ \Theta_1, \dots, \Theta_{N(\Theta,r)} \}$, the corresponding balls.
		We may then use the union bound on the LHS of~\eqref{eq:lemma_covering} and take the supremum inside each ball $\Theta_j$.
		By the triangle inequality, for any $\theta\in\Theta$, $\forall j$, and with probability at least $1-\mdel{4}(\mu)$:
		\begin{align}
		\MoveEqLeft[2]
		\abs{L_{\emp}(\omega_\theta)-L(\omega_\theta)} \nonumber\\
		&\leq \abs{L_{\emp}(\omega_\theta)-L_{\emp}(\omega_{\theta_j})} + \abs{L_{\emp}(\omega_{\theta_j}) - L(\omega_{\theta_j})} \nonumber\\
		&\quad + \abs{L(\omega_{\theta_j}) - L(\omega_\theta)} \nonumber\\
		&\leq \frac{Br}{\tau} +\mu +\frac{Br}{\tau},
		\end{align}
		where the last step follows from Lipschitz continuity of $\log(\cdot)$ and $\omega_{\theta}$ according to Assumption~\ref{Assum:NN_Lip}, and using Proposition~\ref{Prop:convergence_L}.
		Finally, choosing $r=\frac{\tau\mu}{2B}$ concludes the proof.
	\end{proof}
\end{lemma}

The following proposition shows the convergence of $L(\omega_{\hat\theta})$.

\begin{proposition}\label{Prop:L_Ls}
	Let Assumptions~\ref{Assum:knn} and \ref{Assum:NN_Lip} hold. Given $\epsilon>0$, there exists an integer $n_1$ such that, for $n>n_1$,
	\begin{align}
	\pr{L(\omega_{\hat\theta})- L^*\geq \epsilon}\leq \mdel{4}\!\left(\frac{\epsilon}{8}\right).
	\end{align}
	\begin{proof}
		From Lemma~\ref{Lemma:convergence_Sup_L}, with probability at least $1-\mdel{4}(\mu)$, we have that
		\begin{align}
		\abs{L_{\emp}(\omega_{\tilde\theta})-L(\omega_{\tilde\theta})}\leq 2\mu\
		\textnormal{ and }\
		\abs{L_{\emp}(\omega_{\hat\theta})-L(\omega_{\hat\theta})}\leq 2\mu.\label{eq:L2b_conv}
		\end{align}
		Since $\hat\theta$ minimizes $L_{\emp}(\omega_\theta)$, by choosing $\mu=\frac{\epsilon}{8}$, we obtain:
		\begin{align}
		L(\omega_{\hat\theta})&\leq L_{\emp}(\omega_{\hat\theta})+\frac{\epsilon}{4}\leq L_{\emp}(\omega_{\tilde\theta})+\frac{\epsilon}{4}\nonumber\\
		&\stackrel{(a)}{\leq} L(\omega_{\tilde\theta}) + \frac{\epsilon}{2} \stackrel{(b)}{\leq} L^* + \epsilon,\label{eq:L_vicin}
		\end{align}
		where the steps $(a)$ and $(b)$ are due to \eqref{eq:L2b_conv} and Lemma~\ref{Lemma:functiona_apprx}, respectively.
	\end{proof}
\end{proposition}

Proposition~\ref{Prop:L_Ls} implies that $L(\omega_{\hat\theta})$ is close to $L^*$ and, due to the strong convexity of the cross-entropy loss, it can be shown that $\omega_{\hat{\theta}}$ is close to $\omega^*$ (in $\ell_1$ norm).
We continue the proof of the Theorem by defining the following terms:  
\begin{align}
I_{\DV}^{\hat\theta} &\coloneqq \ex_{p(x,y,z)} \big[ \log \hat\Gamma(X,Y,Z) \big]\nonumber\\
&\hspace{1.7cm}- \log \ex_{p(x|z)p(y,z)} \big[ \hat\Gamma(X,Y,Z) \big], \nonumber\\
I_{\NWJ}^{\hat\theta} &\coloneqq 1 + \ex_{p(x,y,z)} \big[ \log \hat\Gamma(X,Y,Z) \big] \nonumber\\
&\hspace{1.7cm}- \ex_{p(x|z)p(y,z)} \big[ \hat\Gamma(X,Y,Z) \big], \nonumber\\
I_{\LDR}^{\hat\theta}&\coloneqq \ex_{p(x,y,z)} \big[ \log \hat\Gamma(X,Y,Z) \big],
\label{eq:expected_I_theta_hat}
\end{align}
where $\hat\Gamma(\cdot)$ is defined in \eqref{eq:Gamma_hat}.
Then from the triangle inequality, we have that
\begin{align}\label{eq:convergence_split}
\Big| \hat I_{\mathtt{est}}^{n,\hat\theta} - I(X;Y|Z) \Big| &\leq \Big| \hat I_{\mathtt{est}}^{n,\hat\theta} - I_{\mathtt{est}}^{\hat\theta} \Big| + \Big| I_{\mathtt{est}}^{\hat\theta}- I(X;Y|Z) \Big|,
\end{align}
where `$\mathtt{est}$' can be replaced with `$\DV$', `$\NWJ$', or `$\LDR$'.
In the following, we show high-confidence convergence of the first and second terms on the RHS of the inequalities~\eqref{eq:convergence_split} due to Lemma~\ref{Lemma:conv_I} and Lemma~\ref{Lemma:conv_I_2}, respectively.

\begin{lemma}\label{Lemma:conv_I}
	Let Assumption~\ref{Assum:knn} hold. For any $\epsilon^*>0$, there exists $n_2$ such that for $n>n_2$ the following bounds hold: 
	\begin{align}
	\pr{ \Big| \hat I_{\DV}^{n,\hat\theta} - I_{\DV}^{\hat\theta} \Big| \geq \frac{\epsilon^*}{2}}&\leq \mdel{5}(\epsilon^*), \label{eq:conv_I_DV}\\
	\pr{ \Big| \hat I_{\NWJ}^{n,\hat\theta} - I_{\NWJ}^{\hat\theta} \Big| \geq \frac{\epsilon^*}{2}}&\leq \mdel{6}(\epsilon^*), \label{eq:conv_I_NWJ}\\
	\pr{ \Big| \hat I_{\LDR}^{n,\hat\theta} - I_{\LDR}^{\hat\theta} \Big| \geq \frac{\epsilon^*}{2}}&\leq \mdel{7}(\epsilon^*), \label{eq:conv_I_LDR}
	\end{align}
	where $\mdel{5}$, $\mdel{6}$, and $\mdel{7}$ are defined in Table~\ref{tab:param}.
	\begin{proof}
		See Appendix~\ref{App:proof_Lemma_conv_I}.
	\end{proof}
\end{lemma}

\begin{lemma}\label{Lemma:conv_I_2}
	Let Assumptions~\ref{Assum:knn} and \ref{Assum:NN_Lip} hold and let $\tau<p_1$. Given $\epsilon^*>0$, there exists an integer $n_1$ such that for $n>n_1$
	\begin{align}
	\pr{ \Big| I_{\mathtt{est}}^{\hat\theta} - I(X;Y|Z) \Big| \geq \frac{\epsilon^*}{2}} &\leq\mdel{4}\left(\frac{\epsilon}{8}\right),\label{eq:conv_I_3}	
	\end{align}
	where `$\mathtt{est}$' can be replaced with `$\DV$', `$\NWJ$', or `$\LDR$', and $\epsilon$ and $\mdel{4}$ are defined in Table~\ref{tab:param}.
	\begin{proof}
		See Appendix~\ref{App:proof_Lemma_conv_I_2}.
	\end{proof}
\end{lemma}

Therefore, if Assumption~\ref{Assum:b} also holds, we may combine Lemma~\ref{Lemma:conv_I} and Lemma~\ref{Lemma:conv_I_2} to yield a high-confidence bound for~\eqref{eq:convergence_split}, which concludes the proof of the Theorem.
\hfill$\square$

%% file: Appendix_F.tex
\section{Proof of Lemma~\ref{Lemma:conv_I}}
\label{App:proof_Lemma_conv_I}

Using the definitions~\eqref{eq:est_CMI} and~\eqref{eq:expected_I_theta_hat}, and the triangle inequality, we have that
\begin{align}
\Big| \hat I_{\DV}^{n,\hat\theta} - I_{\DV}^{\hat\theta} \Big| &\leq \Delta_1 + \myabs{\log\frac{1}{b'}\sum\limits_{\mathcal{B}_\textnormal{prod}^{b'}} \hat\Gamma(x,y,z) \nonumber\\
&\hspace{1.5cm}- \log \ex_{p(x|z)p(y,z)} \big[ \hat\Gamma(X,Y,Z) \big]}\nonumber\\
&\leq \Delta_1 + \frac{p_1}{1-p_1} \frac{1-\tau}{\tau} \Delta_2,\label{eq:DV_diff}
\end{align}
where
\begin{align*}
\Delta_1 &\coloneqq \myabs{\frac{1}{b}\sum\limits_{\mathcal{B}_\textnormal{joint}^b} \log \hat\Gamma(x,y,z) - \ex_{p(x,y,z)} \big[ \log \hat\Gamma(X,Y,Z) \big]},\\
\Delta_2 &\coloneqq \myabs{\frac{1}{b'}\sum\limits_{\mathcal{B}_\textnormal{prod}^{b'}} \hat\Gamma(x,y,z) -  \ex_{p(x|z)p(y,z)} \big[ \hat\Gamma(X,Y,Z) \big]},
\end{align*}
and the last step in \eqref{eq:DV_diff} follows since $\log(\cdot)$ is Lipschitz continuous given that $\hat\Gamma(\cdot)$ is bounded as below by definition,
\begin{align*}
\frac{1-p_1}{p_1}\frac{\tau}{1-\tau} \leq
\hat\Gamma(x,y,z) \leq
\frac{1-p_1}{p_1}\frac{1-\tau}{\tau}.
\end{align*}
Similarly, for the NWJ estimator, we have that
\begin{align}
\MoveEqLeft[.75]
\Big| \hat I_{\NWJ}^{n,\hat\theta} - I_{\NWJ}^{\hat\theta} \Big| \leq \Delta_1 + \Delta_2.\label{eq:NWJ_diff}
\end{align}
Finally, for the last estimator,
\begin{align}
\MoveEqLeft[0.5]
\Big| \hat I_{\LDR}^{n,\hat\theta} - I_{\LDR}^{\hat\theta} \Big| =\Delta_1.
\label{eq:LDR_diff}
\end{align}

We then use Hoeffding's inequality to bound the first terms on the RHS of~\eqref{eq:DV_diff}, \eqref{eq:NWJ_diff}, and~\eqref{eq:LDR_diff}, which results in:
\begin{equation*}
\mathds{P}\big(\Delta_1 \geq \mu\big)\leq 2\exp\Bigg(-\frac{b \mu^2}{2(\log\frac{1-\tau}{\tau})^2}\Bigg).
\end{equation*}
On the other hand, to show the concentration of the second terms on the RHS of~\eqref{eq:DV_diff} and~\eqref{eq:NWJ_diff}, we leverage Theorem~\ref{th:convergence} with \sm{$c=\frac{1-p_1}{p_1}\frac{1-2\tau}{\tau(1-\tau)}$} and $M=\frac{1-p_1}{p_1}\frac{1-\tau}{\tau}$.
Therefore, there exists an integer $n_2$ such that for all $n>n_2$,
\[
\mathds{P}\big( \Delta_2\geq 3\mu \big) \leq \mdel{1}(\mu,c,M ).
\]
From~\eqref{eq:DV_diff}, choosing $\mu=\frac{(1-p_1)\epsilon^*\tau}{2\tau+6p_1-8p_1\tau}$ yields \eqref{eq:conv_I_DV},
while for \eqref{eq:NWJ_diff}, we can choose $\mu=\frac{\epsilon^*}{8}$ to obtain \eqref{eq:conv_I_NWJ}.
Finally for \eqref{eq:LDR_diff}, we choose $\mu=\frac{\epsilon^*}{2}$ to obtain \eqref{eq:conv_I_LDR},
and the proof of Lemma~\ref{Lemma:conv_I} is completed.
\hfill$\square$

%% file: Appendix_G.tex
\section{Proof of Lemma~\ref{Lemma:conv_I_2}}
\label{App:proof_Lemma_conv_I_2}

To express the similarity between $\omega_{\hat\theta}$ and $\omega^*$, we review a lemma from~\cite{mukherjee2019ccmi_conf}, which is based on the strong convexity of the cross-entropy loss and Assumption~\ref{Assum:prob}.

\begin{lemma}(\cite[Lemma~6]{mukherjee2019ccmi_conf})
	\label{Lemma:norm_omega}
	Let Assumption~\ref{Assum:prob} hold.
	Given $\epsilon>0$, if $L(\omega_\theta)\leq L^*+\epsilon$ for some $\theta\in\Theta$, then
	\[
	 \int \abs{\omega^*(x,y,z)-\omega_\theta(x,y,z)}dx\,dy\,dz\leq\eta,
	\]
	where 
    $\eta\coloneqq(1-\tau)\sqrt{2\lambda(\mathcal{X})\epsilon/\alpha},$
	and $\alpha$ is defined in Assumption~\ref{Assum:prob} as the lower bound for the values of the joint and product density functions.
\end{lemma}

From Proposition~\ref{Prop:L_Ls}, we know that, for any $\epsilon>0$, with probability at least $1-\mdel{4}(\epsilon/8)$ 
\[
L(\omega_{\hat{\theta}})\leq L^*+\epsilon. 
\]
Therefore, jointly with Assumption~\ref{Assum:prob}, the requirements of Lemma~\ref{Lemma:norm_omega} are fulfilled.
Let us further define $\Delta_{\omega}(x,y,z) \coloneqq \abs{\omega^*(x,y,z)-\omega_{\hat\theta}(x,y,z)}$, which leads to
\begin{align}
\bar{\Delta}_\omega &\coloneqq \ex_{p(x,y,z)}\Big[\Delta_{\omega}(X,Y,Z)\Big]\nonumber\\
&\,= \int p(x,y,z)\, \Delta_{\omega}(x,y,z)\, dx\, dy\, dz\leq \eta\beta,
\end{align}
and similarly
\begin{align}
\bar{\Delta}'_\omega \coloneqq \ex_{p(x|z)p(y,z)}\Big[\Delta_\omega(X,Y,Z)\Big]\leq \eta\beta.
\end{align}

Next note that, from the continuity of $\Gamma^*(\cdot)$ and $\hat\Gamma(\cdot)$, defined in~\eqref{eq:optim_w} and~\eqref{eq:Gamma_hat}, respectively, we have that:
\begin{align}
\bar{\Delta}_\Gamma &\coloneqq \ex_{p(x,y,z)}\abs{\log\Gamma^*(X,Y,Z) - \log\hat\Gamma(X,Y,Z)}\nonumber\\
&\,\leq \frac{1}{\tau(1-\tau)}\,\bar{\Delta}_\omega, \nonumber\\
\bar{\Delta}'_\Gamma &\coloneqq \ex_{p(x|z)p(y,z)}\abs{\Gamma^*(X,Y,Z) - \hat\Gamma(X,Y,Z)}\nonumber\\
&\,\leq \frac{1-p_1}{p_1}\frac{1}{\tau^2}\,\bar{\Delta}'_\omega. \label{eq:lip_Gamma}
\end{align}
So from the triangle inequality we have:
\begin{align}
&\abs{I_{\DV}^{\hat\theta} - I(X;Y|Z)}\nonumber\\
&\quad \leq \abs{\ex_{p(x,y,z)} \big[ \log\Gamma^*(X,Y,Z)-\log\hat\Gamma(X,Y,Z) \big]} \nonumber\\
&\qquad + \Big\lvert\log \ex_{p(x|z)p(y,z)} \big[\Gamma^*(X,Y,Z) \big]\nonumber\\
&\hspace{3cm}-\log \ex_{p(x|z)p(y,z)} \big[ \hat\Gamma(X,Y,Z) \big] \Big\rvert\nonumber\displaybreak[2]\\
&\quad \stackrel{(a)}{\leq} \abs{\ex_{p(x,y,z)} \big[ \log\Gamma^*(X,Y,Z)-\log\hat\Gamma(X,Y,Z) \big]} \nonumber\\
&\qquad + \frac{p_1(1-\tau)}{(1-p_1)\tau} \abs{\ex_{p(x|z)p(y,z)} \big[ \Gamma^*(X,Y,Z)-\hat\Gamma(X,Y,Z) \big]}\nonumber\displaybreak[2]\\
&\quad\leq  \bar{\Delta}_\Gamma + \frac{p_1(1-\tau)}{(1-p_1)\tau}\,\bar{\Delta}'_\Gamma\, \stackrel{(b)}{\leq}\, \frac{1}{\tau(1-\tau)}\, \bar{\Delta}_\omega + \frac{1-\tau}{\tau^3}\, \bar{\Delta}'_\omega  \nonumber\\
&\quad  \leq \frac{\beta\eta(2\tau^2-2\tau+1)}{\tau^3(1-\tau)},\label{eq:I_DV}
\end{align}
where $(a)$ and $(b)$ are due to Lipschitz continuity of $\log(\cdot)$ and~\eqref{eq:lip_Gamma}, respectively.
Similarly, for the NWJ estimator: 
\begin{align}
&\abs{I_{\NWJ}^{\hat\theta} - I(X;Y|Z)}\nonumber\\
&\quad \leq \abs{\ex_{p(x,y,z)} \big[ \log\Gamma^*(X,Y,Z)-\log\hat\Gamma(X,Y,Z) \big]} \nonumber\\
&\qquad + \abs{\ex_{p(x|z)p(y,z)} \big[ \Gamma^*(X,Y,Z)-\hat\Gamma(X,Y,Z) \big]}\nonumber\displaybreak[2]\\
&\quad\leq  \bar{\Delta}_\Gamma + \bar{\Delta}'_\Gamma \,\leq\, \frac{1}{\tau(1-\tau)}\, \bar{\Delta}_\omega + \frac{1-p_1}{p_1\tau^2}\, \bar{\Delta}'_\omega \nonumber\\
&\quad  \leq \frac{\beta\eta(1+2p_1\tau-p_1-\tau)}{p_1\tau^2(1-\tau)}.\label{eq:I_NWJ}
\end{align}
Finally, for the LDR estimator, we have:
\begin{align}
&\abs{I_{\LDR}^{\hat\theta} - I(X;Y|Z)}\nonumber\\
&\quad = \abs{\ex_{p(x,y,z)} \big[ \log\Gamma^*(X,Y,Z)-\log\hat\Gamma(X,Y,Z) \big]} \nonumber\\
&\quad\leq  \bar\Delta_\Gamma \,\leq\, \frac{1}{\tau(1-\tau)}\, \bar{\Delta}_\omega \,\leq\, \frac{\beta\eta}{\tau(1-\tau)}.\label{eq:I_LDR}
\end{align}

Note that for $\tau<p_1$ we have the following:
$$\frac{1}{\tau(1-\tau)}\leq\frac{1+2p_1\tau - p_1 - \tau}{p_1\,\tau^2(1-\tau)}\leq \frac{2\tau^2-2\tau+1}{\tau^3(1-\tau)}.$$
So by choosing $\eta=\frac{\tau^3(1-\tau)}{(2\tau^2-2\tau+1)\beta}\frac{\epsilon^*}{2}$, $\epsilon$ can be determined from $\eta$ as defined in Lemma~\ref{Lemma:norm_omega}.
This, together with the bounds~\eqref{eq:I_DV}, \eqref{eq:I_NWJ}, and~\eqref{eq:I_LDR}, concludes the proof of Lemma~\ref{Lemma:conv_I_2}.
\hfill$\square$